\documentclass[epj]{svjour}

\usepackage{graphicx}
\usepackage{aas_macros}
\usepackage{amssymb,amsmath}
\def\msun{M_\odot}
\def\mdens{{\rm g~cm^{-3}}}

\def\bdens{{\rm fm^{-3}}}
\def\pne{P_{\rm NE}}

\def\xn{x_{\rm N}}
\def\rhoe{\rho_{\rm E}}
\def\rhon{\rho_{\rm N}}
\def\Mb{M_{\rm b}}
\def\pc{P_{\rm c}}

\begin{document}
\title{Rotating neutron stars with exotic cores: masses, radii, stability}
\author{P. Haensel, M. Bejger, M. Fortin, L. Zdunik}  
\institute{N. Copernicus Astronomical Center, Polish Academy of Sciences, Bartycka 18, 00-716 Warszawa, Poland}
\date{Received: date / Revised version: date}
%
%
\abstract{A set of theoretical mass-radius relations for rigidly
rotating neutron stars with exotic cores, obtained in various
theories of dense matter, is reviewed. Two basic observational
constraints are used: the largest measured rotation frequency (716 Hz)
and the maximum measured mass ($2\;M_\odot$). Present status of
measuring the radii of neutron stars is described. The theory of rigidly
rotating stars in general relativity is reviewed and limitations of
the slow rotation approximation are pointed out. Mass-radius relations
for rotating neutron stars with hyperon and quark cores are
illustrated using several models. Problems related to the
non-uniqueness of the crust-core matching are mentioned. Limits on
rigid rotation resulting from the mass-shedding instability and
the instability with respect to the axisymmetric perturbations are
summarized. The problem of instabilities and of the back-bending
phenomenon are discussed in detail. Metastability and instability of  a
neutron star core in the case of a first-order phase transition, both
between pure phases, and into a mixed-phase  state, are reviewed.
The case of two disjoint families (branches) of rotating neutron
stars is discussed and generic features of neutron-star families and
of core-quakes triggered by the instabilities are considered.
} 
\maketitle
\section{Introduction}
\label{intro}
The determination of radii  ${R}^{\rm obs}_i$ of neutron stars (NS) of
known masses ${M}^{\rm obs}_i$, $(i=1,2,\dots)$ would allow us to
unveil the equation of state (EOS) of neutron-star cores of
density significantly higher than normal nuclear density
$\rho_0=2.7\times 10^{14}\;\mdens$ (corresponding to baryon number
density $n_0=0.16\;\bdens$). To be useful, however,  the
uncertainties in the values  ${M}^{\rm obs}_i,{R}^{\rm obs}_i$
should be sufficiently small (at the level of a few percent) and the
maximum $M^{\rm obs}_i$ should be close to an (unknown) maximum
allowable mass of NS. For the time being, a set of unevenly spaced
${M}^{\rm obs}_i$ was determined\footnote{ see {\tt http://stellarcollapse.org/nsmasses}} \cite{Lat12}, with a maximum value $M^{\rm
obs}_{\rm max}=2.01\pm 0.04\;\msun$ \cite{Antoniadis2013} being a very strong constraint
 on the EOS. The precise measurement of $R$ is
still a challenge for observers (see sect.~\ref{Obs}).

For $\rho<\rho_0$ constituents of matter are well established:
nucleons and electrons, with a small admixture of muons at the upper
subnuclear density segment where the Fermi energy of electrons
exceeds the muon rest energy. However, we expect that the central
density of neutron stars with $M>1\;\msun$ is larger than
$1.5\rho_0 - 2\rho_0$. For $M=2\;\msun$ the star's central density  may be
as high as $\sim 7\rho_0$. For $\rho\gtrsim 2\rho_0$ even the actual
hadronic constituents of the NS core are uncertain: are they just
nucleons (zero strangeness), or more generally baryons, {\it i.e.}
nucleons and hyperons (non-zero strangeness)? Maybe the density
 realized there is sufficient for a phase transition to  quark matter?
Finally, maybe in addition to baryons, real kaons or pions forming a
boson condensate are present there? The uncertainty grows with
increasing $\rho$, which is expected to be as high as $8\rho_0 -
10\rho_0$ at the center of the most massive NS. This uncertainty results
from the lack of precise knowledge of strong interactions and the
approximations (often uncontrollable) of the many-body theories of
super-dense hadronic matter. The uncertainty in the structure and
composition of super-dense matter implies an even larger uncertainty
in the EOS of NS cores.

All neutron stars rotate and there are many millisecond pulsars
(MSP) with rotation frequency $f>500~$Hz (10 accreting
X-ray pulsars and 14 radio/gamma-ray pulsars), and a radio MSP with
716 Hz is observed (sect.~\ref{Obs}). In the present paper we restrict 
ourselves to rigid rotation and axisymmetric approximation for rotating 
NS models, nevertheless for the completeness sake we briefly recall the 
general picture. The approximation of rigid, axisymmetric rotation holds
extremely well already a few minutes after the NS is born in the supernova core
collapse.  Prior to that one expects a differentially-rotating, hot and
lepton-rich proto-NS with high entropy, which cannot be properly described by a
cold, catalyzed matter EOS. During this period dynamo mechanism and convection may operate,
increasing the interior magnetic field and leading to magneto-hydrodynamical instabilities
\cite{Ciolfi2014}. Shortly after the birth, rigid rotation sets in 
due to the presence of viscosity. However, for sufficiently high rotation rates,
parametrized by the kinetic energy $T$ to potential energy $W$ ratio,
$\beta=T/\lvert W\rvert$, a {\it dynamical} triaxial bar-mode instability may
arise in rigidly rotating stars; relativistic calculations indicate 
critical $\beta\simeq 0.24$. Substantial differential rotation facilitates 
the onset of these dynamical
instabilities - they may occur at a low $\beta\approx 0.01$. Another, {\it
secular} bar-mode instability, {\it driven} by the dissipation due to viscosity or
the emission of gravitational waves sets in at $\beta\simeq 0.14$ (for a
review see \cite{AnderssonC2007}). We also expect that accreting NSs may be
prone to the Rossby-type instabilities (r-modes), driven by the Coriolis
force \cite{AnderssonK2001}.

The dependence of the  radius of a NS with an exotic core on its mass and the imprint of rotation on the mass-equatorial radius relation and the stability 
of rotating NS configurations is the main topic of the present paper. In our review we try to present some generic features of rotating NS with exotic (E) cores, 
with E being:  hyperon matter, quark matter, or a baryon phase with a boson (pion or kaon) condensate. NS with  E-core will be compared  with standard nucleon NS models, hoping that the differences between theoretical models, confronted with observations, will help to unveil the true EOS of NS.

In order to study axisymmetric  hydrostatic equilibria of rotating NS in general relativity (GR), approximate, as well as exact numerical methods of solving Einstein's 
equation were developed; we briefly discuss them in sect.~\ref{theory}. 
We discuss the precision which should be reached in the
2D calculations to study the  mass shedding limit, the spin evolution and stability with respect to axisymmetric perturbations  of rotating NS models. 

We consider  a  relativistic star in the perfect fluid approximation. In
general, the metric of space-time around a rotating NS is essentially
different from that around a static star ($f=0$). For $f=0$ the
metric depends only on the NS  mass $M$ and it is the Schwarzschild
metric. For $f>0$ the metric depends explicitly on the matter and
pressure distribution inside a rotating star, which makes it
dependent on the EOS; it also depends on $f$ through the effect of the 
dragging of the inertial frames. In particular, this refers to the
innermost stable orbit around an accreting  MSP (sect.~\ref{theory}).
Some properties of the circular orbits around NS and their relation
to the Keplerian limit are reviewed in sect.~\ref{kepler}.

The EOS of dense matter with a transition from a normal (N) phase to an exotic (E) 
one has some particularities. Generic features of such an EOS are reviewed in sect.~\ref{metastab}. We start with the simplest case of N-E transition in full thermodynamic equilibrium, reviewed in sect.~\ref{pure.mixed}. Then we consider a more complicated case including the possibility of a metastable state and nucleation of the E-phase in the N-one  in sect.~\ref{nucleation}.

The theoretical $M-R$ relation (here $R$ is the equatorial radius) depends 
on the rotation frequency. The region in the  $M-R$ plane allowed for rotating configurations is affected by the presence of an exotic core in massive NS. The
maximum allowable NS mass is a functional of the EOS. Its value for
non-rotating stars, $M_{\rm max}^{\rm stat}$, has to satisfy $M_{\rm
max}^{\rm stat}[{\rm EOS}]>2\;\msun$, the largest observed mass. Rotation increases $M_{\rm max}$
only by a few percent even at $716\;$Hz, but for the minimum mass 
$M_{\rm min}^f$  (which for $f=0$ is $M^{\rm stat}_{\rm min}\approx 0.1\;\msun$,  see
\cite{HPY2007book} and references therein), the effect of rotation
is dramatic and depends indirectly also on the phase transition to an
exotic high density phase (sect.~\ref{exoNS}). This property of $M_{\rm min}^f$ 
can be used to derive the EOS dependent lower bound on the mass 
of the fastest 716 Hz pulsar (sect.~\ref{mass716Hz}). We present in detail
examples of families of rotating NS models with nucleon, hyperon, 
and quark cores, showing the differences between these families.

Limits on the frequency of rotation of NS are reviewed in sect.~\ref{kepler}.  
There is an upper bound for the frequency of rotation for each
given baryon (rest) mass of a NS  $M_{\rm b}$,
corresponding to a non-rotating (static) mass $M_{\rm s}$. It
results  from the mass-shedding instability at the equator and is 
called Keplerian frequency $f_{\rm K}(M_{\rm s})$. This bound is quite  sensitive
 to the EOS, because it depends on the radius of non-rotating
configuration (sect.~\ref{mass-shedding}). There is also a theoretical 
maximum  frequency  for all stably rotating configurations of NS, 
  $f_{\rm max}=1500\;{\rm Hz} - 2000\; {\rm Hz}$ which  depends on the EOS   (sect.~\ref{fmax.stab}). The highest measured frequency of a pulsar which as 
  for today is 716 Hz, results in an EOS dependent  constraint on the mass of this fastest pulsar (sect.~\ref{mass716Hz}).

In sect.~\ref{dynamics} we review various aspects of the spin evolution, dynamics,
and stability of rotating NS with exotic cores. 
The softening of the EOS associated with a transition to an exotic 
phase can lead to a phenomenon of back bending  (spin-up induced by an 
angular momentum loss), reviewed in sect.~\ref{sect:bbstab}. In particular, we 
point out the possibility of the existence of unstable segments of configuration sequences,  splitting the stable back bending fragment of the spin evolution track into two separate branches.

The instability induced by the softening of the EOS due to a  first order phase 
transition into an exotic phase is discussed in sect.~\ref{sect:firstinstab}. A
sufficiently strong  softening of the EOS  can lead to splitting a
single one-parameter family of hydrostatic configurations of NS 
into two separate (disjoint) families. This feature is valid not only for static NS, but also  for rotating NS models with constant $f$. The static criterion for the split into two  branches is valid also for rigidly rotating configurations 
(sect.~\ref{sect:firstinstab}). We then review, in sect.~\ref{sect:minicoll}, the possibility of a minicollapse of a NS due to the nucleation of the E-phase at its center during the NS evolution,  and the role of metastability of the 
N-phase core in this process.  The astrophysical signatures of a mini-collapse are 
described in sect.~\ref{sect:mini-signatures}. 
In sect.~\ref{formation.crust} we review the effect of the crust formation scenario on the 
$M - R$ relation.  

Our conclusions are summarized in the final sect.~\ref{conclusions}.

\section{Observational constraints from spin frequency and radius measurements}
\label{Obs}
The recent discovery of two $2\;\msun$ pulsars \cite{Demorest2010,Antoniadis2013}
provides an important constraint on the poorly known equation of
state at supra-nuclear density. In this section we summarize the current status of measurements of radius and spin frequency of NS.

\subsection{Radius}
The radius of a NS can in principle be extracted from the analysis of X-ray spectra emitted by the NS atmosphere (see \cite{P14} for a review). However even in the case of a non-rotating NS, due to the space-time curvature, only the apparent radius:
\begin{equation}
R_\infty=\frac{R}{\sqrt{1-2GM/Rc^2}}
\end{equation} is constrained by the modelling. It actually depends on \textit{both} the radius and the mass. Measurements are complicated since they depend on the distance to the NS, its magnetic field, the composition of its atmosphere and the interstellar absorption (see {\it e.g.} \cite{HC14}). On the one hand, the magnetic field of isolated NS is likely to be large ($B > 10^9$~G) and thus will affect their spectra, and the chemical composition of their atmosphere is unknown and difficult to determine. On the other hand, NS that undergo periods of accretion of matter from their binary companion are believed to have a low magnetic field (due to accretion-induced decay), and an atmosphere likely to be composed of light elements (H, possibly He \cite{CH13,HC14}). Among such objects one can distinguish quiescent X-ray transients (QXT), NS in a binary system observed when the accretion has stopped or is strongly reduced, and bursting NS (BNS) {\it i.e.} NS from which recurring and very powerful bursts, so-called photospheric radius expansion (PRE) bursts, are observed. These sources are even more promising when they are located in globular clusters whose distance is likely to be accurately measured. $R_\infty$ can also be constrained by the modelling of the shape of the X-ray pulses observed from rotation-powered radio millisecond pulsars (RP-MSP) in particular if their mass is known from radio observations.

\begin{figure}
\resizebox{\hsize}{!}{\includegraphics*{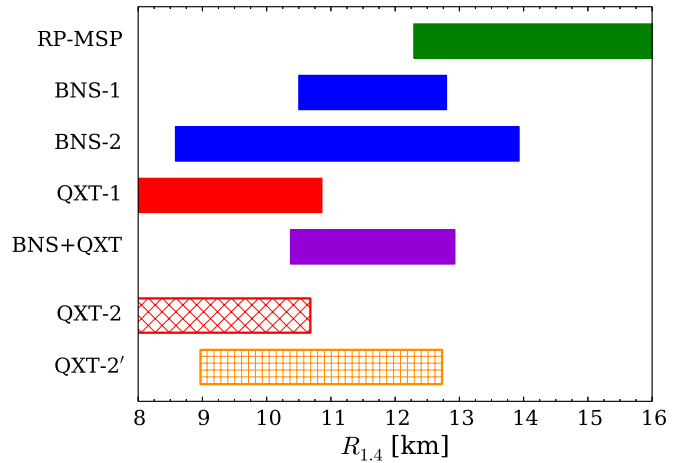}}
\caption{(Colour online) Constraints on the radius $R_{1.4}$ of a $1.4\;\msun$ NS from different types of NS: a RP-MSP \cite{Bogdanov2013}, BNS-1 \cite{Natilla2015}, BNS-2 \cite{Ozel2013}, QXT-1 \cite{Guillot2014}, BNS+QXT \cite{SteinerLattBrown2013}. Constraints QXT-2 and QXT-2$^\prime$ are included for discussion only, see text for details. The constraints correspond to 2-$\sigma$ error bars.}
\label{fig:radcons}
\end{figure}

Fig.~\ref{fig:radcons} shows the most recent constraints on the radius $R_{1.4}$ of a $1.4\;\msun$ NS obtained for various types of sources (see details in \cite{FortinZHB2015Rhyp}). The constraints QXT-1 and RP-MSP being mutually exclusive, so far no consensus on $R_{1.4}$ can be reached.
However, the determination of the radius of a NS is subject to many assumptions,
uncertainties and systematics effects, (see {\it e.g.} table 1 in \cite{P14}). Obtaining constraints from the PRE bursts of BNS is still subject to uncertainties and debates  in particular concerning the modelling of the phenomenon itself, the selection of bursts to be used (hard state X-ray bursts {\it vs.} soft state ones) and the composition of the atmosphere (see \cite{Natilla2015,Poutanen2014,SteinerLattBrown2013,Ozel2013,OP15a,OP15b}). As far as QXT in globular clusters are concerned, the composition of the atmosphere and the amount of interstellar absorption, quantified by the so-called `equivalent hydrogen column density' $N_{\rm H}$, are unknown and significantly affect results \cite{GS13,LS14,HC14}. For example among the five QXT studied in \cite{GS13} (constraint QXT-2 in fig.~\ref{fig:radcons}), one of them: NGC 6397, has a substantially smaller $R_\infty$ compared to the value obtained for the four other sources. As a consequence the constraint QXT-2 that is derived from these five sources suggests a small NS radius. However, while for one of the five sources observations suggest a hydrogen composition for the atmosphere: $\omega\,$Cen \cite{HC04}, the composition of the atmosphere of the four other sources, including NGC 6397, is still unknown. Using a helium atmosphere instead a hydrogen one, a larger $R_\infty$ is obtained for NGC 6397 \cite{HC14}. The constraint QXT-2$^\prime$ corresponds to the QXT-2 one when NGC 6397 is not included: it then favours larger radii. The quantity $N_{\rm H}$ can in principle be constrained thanks to observations in various wavelengths or derived when fitting the X-ray spectra. Large discrepancies between the values derived for $N_{\rm H}$ using these two approaches are however observed and as a consequence the constraint on $R_\infty$ can vary by as much as a factor 2 (for $\omega\,$Cen in \cite{GS13}). Finally, the uncertainty on the distance to globular clusters can be as large as 25\% \cite{LS14} and further affects the constraint on the radii (see {\it e.g.} \cite{GS13,LS14,HC14}).
Last but not least, taking into account NS rotation strongly complicates the analysis of the collected X-ray spectra. Both QXT and BNS are likely to rotate at a frequency of few hundreds of Hz which is expected to affect the radius determination by $\sim10\%$ according to \cite{Poutanen2014,BO14}.

Due to uncertainties in both the observations and the modelling of QXT, BNS and RP-MSP, no stringent constraint on the radius of NS can currently be derived. However the next generation of X-ray telescopes such as NICER \cite{NICER}, Athena \cite{Athena+} and, possibly, a LOFT-like \cite{LOFT} missions promise measurements of radii with an accuracy of few percents. Together with a constraint on the maximum observed NS mass, a simultaneous measurement of NS mass and radius with such a precision could enable to constrain the NS EOS.

\subsection{Spin frequency}
Since the discovery of the first pulsar in 1967 \cite{HB68}, later identified as being a NS \cite{HB68,G69}, $\sim$ 3000 NS have been observed in all wavelengths, most of them as radio pulsars (see \cite{K10} for a review). Among those one can distinguish two populations: the so-called `normal pulsars' with periods of the order of few seconds and the `millisecond pulsars' (MSP) which as their name indicate have a period of the order of few milliseconds \cite{L08}. These are believed to be old NS that have been `recycled' {\it i.e.} spun-up to millisecond periods by the accretion of matter from a binary companion \cite{Alpar1982,Radhakrishnan1982}.

During the recycling process, a binary system can be observed as an X-ray source and its pulsar as an X-ray millisecond pulsar (XMSP). The spin frequency can be determined or estimated for three different types of XMSP \cite{WK08}:
\begin{itemize}
\item accreting X-ray millisecond pulsars (AXMSP): X-ray pulsations due the presence of hotspots at the surface of the rotating neutron star have been observed from these sources. The spin frequency of 15 AXMSP has been measured with a great accuracy (see \cite{PW12} for a review). 
\item nuclear X-ray millisecond pulsars (NXMSP): they exhibit oscillations during thermonuclear X-ray bursts. The frequency of the oscillations is thought to be at or close to the pulsar spin frequency, though there are still some uncertainties on the physical process that triggers the oscillations \cite{W12}. Therefore for these sources, the measurement of the spin frequency is indirect and has an uncertainty of few hertz. The spin frequency of 10 NXMSP has been determined so far;
\item twin kilohertz quasi-periodic oscillations have been observed in several systems. However, their interpretation and the precise link with the rotation of the neutron star is still unclear (see {\it e.g.} \cite{W12}).
\end{itemize}

\begin{figure}
\resizebox{\hsize}{!}{\includegraphics*{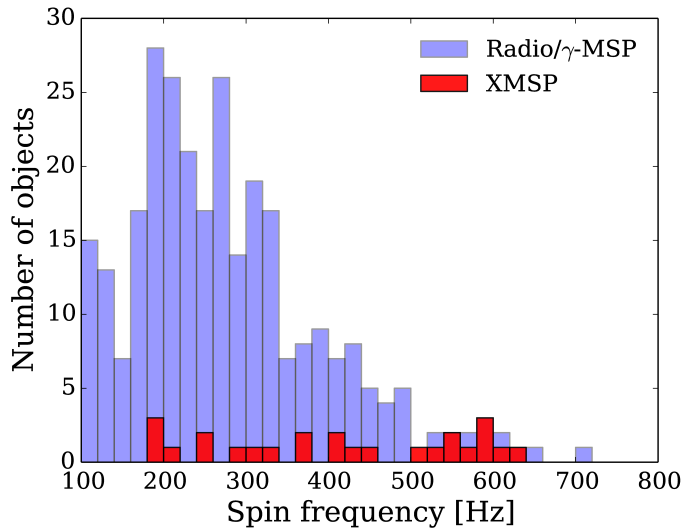}}
\caption{(Colour online) Histogram of the spin frequency of radio/gamma-ray MSP and XMSPs.}
\label{fig:histf}
\end{figure}

Fig.~\ref{fig:histf} shows the frequencies of currently observed radio and gamma-ray pulsars (data from the ATNF Pulsar
Catalogue\footnote{\tt http://www.atnf.csiro.au/people/pulsar/psrcat} \cite{ATNF}) and XMSP rotating at a frequency larger than $100$~Hz. Out of $\sim 2500$ radio and gamma-ray pulsars with measured period, $11\%$ of them have a spin frequency larger than 100 Hz. So far, the fastest rotating XMSP is 4U1608$-$522 with $f=620$~Hz \cite{GM08} and the fastest rotating MSP is PSR J1748$-$2446a in the globular cluster Terzan 5 with $f=716$~Hz \cite{HR06}. Oscillations at a frequency of 1122 Hz in one type I X-ray burst of XTE J1739$-$285 were reported \cite{KP07} but not observed later \cite{GM08}.

Although with current observational techniques submillisecond pulsars could in principle be detected, so far all attempts were unsuccessful (see {\it e.g.} \cite{D00,P10,DE11}). Thus this might indicate the existence of a mechanism that prevents accreting NS from reaching submillisecond periods. The interaction between the NS magnetic field and the accretion disk (see {\it e.g.} \cite{AG05}) or the loss of angular momentum due to the emission of gravitational radiation (see {\it e.g.} \cite{B98,WK08,HM11}) may inhibit the recycling process and thus the formation of submillisecond pulsars.
The consequences of the existence of the fastest rotating pulsar with $f=716$~Hz are discussed in sect.~\ref{kepler}.

\section{Rotating stars in general relativity}
\label{theory}
After a few decades since the discovery of the spherically-symmetric, static solution for matter
distribution by Tolman, Oppenheimer and Volkoff
(\cite{Tolman1939,OppenheimerV1939}) - the TOV equations - the interest of 
researchers on rotating, relativistic stars was revived in the 1960s  
Golden Era of GR. The breakthrough came just in time for the discovery of the first pulsar 
with the work of Hartle \cite{Hartle1967}, who devised a {\it slow-rotation
approximation} to an exact solution by treating rigid rotation as a small perturbation of the
spherically-symmetric TOV background solution. In
quasi-Schwarzschild coordinates $(t,r,\theta,\phi)$, such a stationary,
axisymmetric spacetime is described by a generic metric\footnote{In this section, 
for brevity we adopt the geometric convention $G=c=1$.}  
\begin{eqnarray} 
\mathrm{d}s^2 &=& -H(r,\theta)^2 \mathrm{d}t^2 + Q(r, \theta)^2 \mathrm{d}r^2 \nonumber \\ 
&+& r^2K(r,\theta)^2\left(\mathrm{d}\theta^2 
+ \sin^2\theta\left(\mathrm{d}\phi - \omega(r,\theta)\mathrm{d}t\right)^2\right), 
\label{eq:metric_qSch}
\end{eqnarray}
where the $\omega$ function is the angular velocity of the free-falling observers
(which corresponds to the frame dragging of inertial frames). In the first-order expansion in terms of
the star's angular spin frequency $\Omega$, outside the star the metric function 
$\omega = 2J/r^3$, {\it i.e.}, it is proportional to the total stellar angular momentum $J$. 
With respect to the metric used in the TOV solution, the $O(\Omega)$ metric differs 
only by the $\omega dt$ term, 
\begin{eqnarray}
\mathrm{d}s^2 &=& - e^{2\nu(r)}\mathrm{d}t^2 + \left(1 - \frac{2m(r)}{r}\right)^{-1}\mathrm{d}r^2 \nonumber \\  
&+& r^2\left(\mathrm{d}\theta^2 + \sin^2\theta\left(\mathrm{d}\phi - \omega(r,\theta)\mathrm{d}t\right)^2\right) 
+ O(\Omega^2). 
\label{eq:metric_sr_1stord}
\end{eqnarray}
In addition to the usual TOV ordinary differential equations for $m(r)$, $\nu(r)$ and 
pressure $p(r)$ the off-diagonal $t\phi$ component of Einstein's equations provides 
a differential equation for $\omega(r)$, or equivalently, $J$ (which then can be used 
to define the moment of inertia of the star $I = J/\Omega$). 

Subsequently, Hartle and Thorne \cite{HartleT1968}, and Sedrakyan and Chubaryan
\cite{SedrakyanC1968} obtained a second-order, $O(\Omega^2)$ solution of the
slow-rotation approximation. Within this approximation, the star's angular momentum, 
the fluid velocity and the frame-dragging term are exactly the same as in the $O(\Omega)$ 
order (they are functions of odd powers of $\Omega$). 
What is affected are however the diagonal metric terms and 
pressure and energy density distributions. The metric reads 
\begin{eqnarray}
\mathrm{d}s^2 &=& - e^{2\nu(r)}\left(1 + 2\left(h_0 + h_2P_2(\cos\theta) \right)\right)\mathrm{d}t^2 
\nonumber \\ 
&+& \left(1 - \frac{2m(r)}{r}\right)^{-1} 
\left( 1 + \frac{2\left(m_0 + m_2P_2(\cos\theta)\right)}{r - 2m(r)} \right)\mathrm{d}r^2 \nonumber \\  
&+& r^2\left(1 + 2\left(v_2 - h_2\right)P_2(\cos\theta)\right) \times \nonumber \\ 
&& \left(\mathrm{d}\theta^2 + \sin^2\theta\left(\mathrm{d}\phi - \omega(r,\theta)\mathrm{d}t\right)^2\right) 
+ O(\Omega^3),  
\label{eq:metric_sr_2ndord}
\end{eqnarray}
where $h_0(r)$ and $m_0(r)$ describe the monopole deformations, 
and $h_2(r)$, $m_2(r)$ and $v_2(r)$ describe the quadrupole deformations (the dipole 
term is identically zero). The function $P_2(\cos\theta)$ is the Legendre 
polynomial. The pressure inside the star is modified as follows: 
\begin{equation} 
p(r,\theta) = p(r) + (\rho + p)\left(p_0 + p_2P_2(\cos\theta)\right),
\label{eq:pressure_sr_2ndord}
\end{equation} 
with $\rho$ denoting the energy density. Similarly to the first-order
expansion, the solution consists of the TOV background solution, supplemented
with an additional set of first-order ordinary differential and algebraic
equations for the monopole and quadrupole terms. The gravitational mass $M$  
of a star with angular momentum $J$ is 
\begin{equation} 
M = {\cal M} + m_0({\cal R}) + J^2/{\cal R}^3 + O(\Omega^4), 
\label{eq:mass} 
\end{equation} 
where ${\cal M}$ and ${\cal R}$ are the mass and radius of a non-rotating configuration 
with the same $\rho_c$. Rotational corrections deform the star to a spheroid shape. 
The $\theta$ dependence of the radius is  
\begin{equation} 
R(\theta) = {\cal R} + \xi_0({\cal R}) + \xi_2({\cal R})P_2(\cos\theta), 
\label{eq:spheroid}
\end{equation} 
with $\xi_0$ and $\xi_2$ being functions of $p_0(r)$, $p_2(r)$, $v_2(r)$, $h_2(r)$, as well 
as the equation of state and structure. The equatorial circumferential radius equals  
\begin{equation} 
R^{circ}_{e} = \sqrt{g_{\phi\phi}(R_e,\theta=\pi/2)}, 
\label{eq:ht_rcirc}
\end{equation} 
for which one can evaluate the exterior solution of $g_{\phi\phi}$ at the surface; 
$R_e$ is the coordinate equatorial radius obtained by integrating the equations 
of structure. 

A second-order slow-rotation approximation allows for defining the star's
quadrupole moment $Q$ by comparison of the metric terms with their Newtonian
analogues. Consequently it allows to characterize the exterior metric of a
rotating object using its multipole moments: gravitational mass $M$, angular
momentum $J$ and quadrupole moment $Q$. This, together with the fact that
slow-rotation approximation offers a direct and intuitive link to a
spherically-symmetric Schwarzschild configurations (by using the same system of
coordinates), as well as a relative simplicity of the set of ordinary
differential equations to solve is its biggest advantage. 

There are drawbacks of this approach, however. Slow-rotation approximation
cannot be applied to all rotation rate: by definition $\Omega/\Omega_{\rm K} \ll 1$,
where $\Omega_{\rm K}=2\pi f_{\rm K}$ is the {\it Keplerian} (mass-shedding) angular frequency. Moreover,
the very definition of a spheroid shape in eq.~(\ref{eq:spheroid}) prevents the
star to accurately reproduce the mass-shedding limit. 
 Thirdly, $O(\Omega^2)$
definitions of the gravitational mass $M$ and angular momentum $J$ are in
general not accurate enough to robustly indicate an instability (by means, for
example, the arguments based on the turning-point theorem of
\cite{Sorkin1981,Sorkin1982,FriedmanIS1988}, which is a sufficient condition 
for instability). 

To obtain accurate results for any rotation rate, one needs to change the way 
the problem is posed. 
On these grounds highly accurate numerical schemes like \cite{BGSM1993,StergioulasF1995,AnsorgKM2002} 
were developed. Among other things, it is useful to abandon the Schwarzschild coordinates 
in favor of {\it e.g.}, quasi-isotropic coordinates. The metric is then expressed as  
\begin{eqnarray} 
\mathrm{d}s^2 &=& -e^{2\nu(r,\theta)}\mathrm{d}t^2 + e^{2\mu(r,\theta)}\left(\mathrm{d}r^2 
+ r^2\mathrm{d}\theta^2\right)\nonumber \\ 
 &+& e^{2\psi(r,\theta)}\left(\mathrm{d}\phi - \omega \mathrm{d}t\right)^2, 
\label{eq:metric_qi}
\end{eqnarray} 
where the $\nu$ potential is as previously related to the gravitational
potential of the source, and $e^\mu$ and $e^\psi$ are called conformal factors.
Note the different relation between the $r$ and $\theta$ coordinates with
respect to {\it e.g.} eq.~(\ref{eq:metric_sr_1stord}). 
The Einstein equations to be solved can be derived in a number of ways.  
A particularly interesting, widely accepted and successful
approach in numerical relativity is the 3+1 decomposition of spacetime {\it i.e.}, a 
specific `slicing' of the four-dimensional spacetime into spacelike three-dimensional
hypersurfaces in order to deal with the three-dimensional tensor fields to
obtain solutions (see, {\it e.g.}, \cite{Gourguolhon2007,BaumgarteS2010} 
as well as \cite{Gourgoulhon2010} and \cite{FriedmanS2013} for the specialized 
case of rotating relativistic stars). In the quasi-isotropic gauge 
with a choice of slicing (maximal slicing), 
the Einstein equations for a stationary, axisymmetric star are expressed as a
system of four coupled non-linear elliptic partial differential (Poisson-like)
equations: 
\begin{eqnarray} 
\Delta e^\nu = \sigma_1,\quad & \Delta \omega = \sigma_2,\nonumber\\
\Delta e^{\nu\mu} = \sigma_3,\quad & \Delta e^{\nu\psi} = \sigma_4,
\label{eq:equations_3p1}
\end{eqnarray} 
where the right-hand-sides of each equation, $\sigma_i$, are the source terms
describing non-linear metric terms and matter via the energy-momentum tensor, 
usually assumed to describe the perfect fluid. The only
boundary condition for this system exists in spatial infinity and is provided
by the asymptotically flat metric. Global quantities, such as gravitational
mass $M$ are formulated as surface or volume integrals using the asymptotic
behavior of appropriate metric functions far from the source (the so-called ADM
mass, \cite{ADM1959}), or by exploiting the symmetries of the problem (the so-called
Komar mass for stationary spacetimes by taking into account the existence of a time Killing vector, \cite{Komar1959}). Similar reasoning applies to the
angular momentum $J$. Redefining $e^\psi$ as $Br{\rm sin}\theta$, one can get the
relation between the coordinate and the circumferential radius. Analogous to
eq.~(\ref{eq:ht_rcirc}), the circumferential equatorial radius of the star, {\it i.e.},
the length of the equator divided by $2\pi$, is simply 
\begin{equation}
R^{circ}_{\rm e} = B R_{\rm e},
\end{equation} 
with $R_{\rm e}$ being the coordinate equatorial radius. In what follows
we will denote the circumferential equatorial radius of the rotating  star by $R_{\rm eq}$.

\subsection{Accuracy of solutions}
\begin{figure}
\resizebox{\hsize}{!}{
\includegraphics{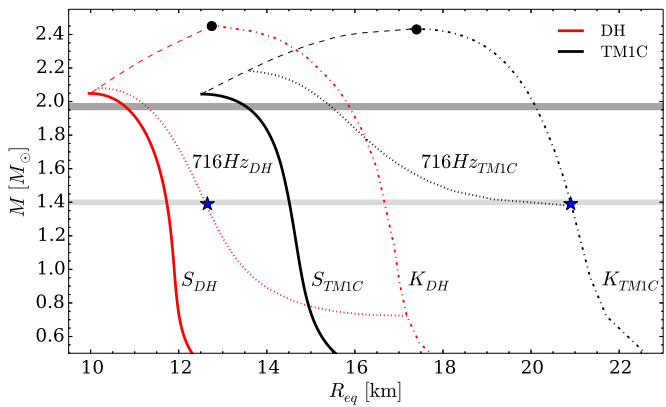}}
\caption{Gravitational mass $M$-equatorial radius $R_{\rm eq}$ plane for rotating
stars.  The figure shows two sets of curves: for the DH EOS
(\cite{DouchinH2001}, red lines) and TM1C EOS (\cite{GusakovHK2014}, black
lines): solid lines marked with the {\bf S} symbol denote static (non-rotating)
sequences, dotted lines are constant spin frequency $716$ Hz sequences,
dash-dotted lines marked with the {\bf K} symbol are the mass-shedding
configurations, and dashed lines mark the sufficient conditions for the
instability with respect to the axisymmetric perturbations. Stars denote the
configurations depicted on fig.~\ref{fig:tm1c_sly4}. The physical reason for 
high stiffness of the TM1C EOS for $M\lesssim 1.4\;\msun$, leading to 
large radii of NS models for this mass range,  is explained in sect.~\ref{sect:hypNS}}.
\label{fig:mr-static-kepler}
\end{figure}
\begin{figure}
\begin{center}
\resizebox{0.95\hsize}{!}{
\includegraphics{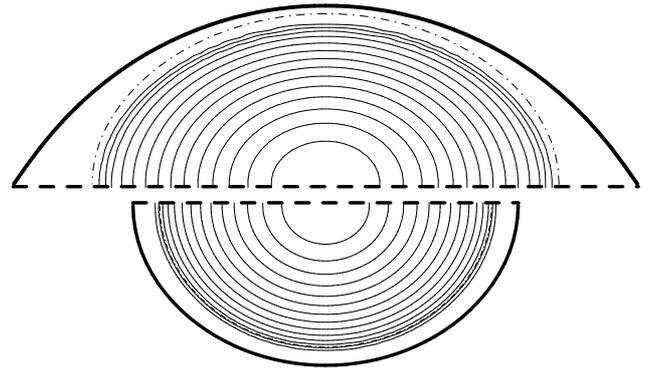}
}
\end{center}
\caption{Non-spherical shapes of NS deformed due to rapid rotation.
Isocontours denote constant fluid proper energy density; vertical direction is aligned 
with star's angular momentum. Thicker line
correspond to the surface. Upper slice corresponds to the TM1C EOS NS 
with $M=1.4\,\msun$, rotating at 716 Hz which for this configuration is the
mass-shedding limit (Keplerian rotation; note the cusp at the equator). Lower
slice corresponds to the DH EOS with the same mass and spin frequency. The configurations 
correspond to stars marked on fig.~\ref{fig:mr-static-kepler}. The
circumferential radius of TM1C NS is 20.9 km, whereas for DH NS it is 12.9 km.
The results were obtained using the {\tt LORENE/nrotstar} code.} 
\label{fig:tm1c_sly4}
\end{figure}

An additional important aspect of solving numerically the Einstein equations
is the choice of numerical methods. In some applications the {\it spectral
methods} prove to be superior over traditionally more widely used finite-differences
methods. With the spectral decomposition, functions are expanded in terms of adequately 
chosen basis functions, and resulting algebraic equations for the expansion
coefficients are solved. When properly implemented, the difference between the
series expansion and the real solution vanishes like $e^{-N}$, where $N$ is the number
of expansion coefficients (the evanescent error). As an example of a real implementation, 
the  formulation  of \cite{BGSM1993} is using spectral methods in a numerical library {\tt
LORENE}\footnote{\tt http://lorene.obspm.fr} in a {\tt nrotstar} code; other
highly-accurate implementations can easily reach machine precision
\cite{AnsorgKM2002}. Note also that numerical relativity with spectral methods is
particularly suitable for precision studies of instabilities, due to very low numerical 
viscosity of spectral methods (see {\it e.g.}, \cite{BejgerHZ2005,ZdunikBHG2006,ZdunikBHG2007,ZdunikBHG2008} 
and sect.~\ref{sect:bbstab} of this article). 

The accuracy of a numerical solution may be checked in a number of ways. For
example, for stationary asymptotically flat spacetime, the Komar mass is in
principle equal to the ADM mass - their difference expresses thus the
imperfection of numerical solution and is proportional to the accuracy
achieved.  A very sensible and widely used accuracy indicator is a relativistic
generalization of the classical virial theorem by
\cite{GourgoulhonB1994a,GourgoulhonB1994b}, applicable in case of
asymptotically-flat four-dimensional spacetimes such as in the case of rotating
relativistic stars. 

For an example, figs.~\ref{fig:mr-static-kepler} and \ref{fig:tm1c_sly4} show the effect 
of rigid rotation on the mass-radius $M(R)$ sequences for two recent EOS - a `standard' 
nucleonic DH EOS of \cite{DouchinH2001} and stiff TM1C EOS that includes hyperons.
 The latter is a non-unified EOS (see discussion in sect.~\ref{sect:hypNS}) where 
the DH EOS is used for the crust and the model by \cite{GusakovHK2014} for the core. 
Configurations were obtained with the use 
of {\tt LORENE/nrotstar}. Specifically, fig.~\ref{fig:tm1c_sly4} 
shows how strongly the shape of a rapidly-rotating relativistic star depends 
on the EOS. Note that the configuration at the verge of a breakup 
(mass-shedding limit, which is related to a cusp on the stellar surface at the equator) 
cannot be accurately simulated with the slow-rotation approximation.   

Apart from the shape, rigid rotation changes the values of global parameters of a
star; it increases its equatorial radius $R_{\rm eq}$ and also the maximum gravitational
mass $M_{\rm max}$ with which the star can still be stable for a given central EOS 
parameters. The maximum increase of $M_{\rm max}$ is, with respect to
the non-rotating configurations, about 15-20\% for dense matter described by 
realistic hadronic EOSs. Note that the mass increase caused by rotation 
cannot be therefore proposed as a general solution to the problem of maximum mass 
{\it decrease} caused by a substantial phase transition/softening in some exotic EOS, 
in order to reconcile their inconsistency with recent observational data.  
The maximal increase of $R_{\rm eq}$ is about 30-40\% for the
mass-shedding configurations (see sect.~\ref{kepler} for more details).
 
Exact numerical solutions were compared with the slow-rotation approximation
in a number of articles. Notably, differences between the results obtained
using the early implementation of {\tt LORENE/nrotstar} \cite{BGSM1993} and the
results obtained by \cite{WeberG1992} are described in \cite{SalgadoBGH1994};
the differences in mass for rotating maximum mass models is reported to be of
about 5\%, whereas the differences in radii about 15\%. The slow-rotation
approximation is particularly sensitive to quantities that depend on the
derivatives of the metric functions, {\it e.g.} the radius or the innermost stable
circular orbit around a rotating compact star. In a comparison performed in
\cite{ZdunikHGG2000}, where  the radii of orbits around strange quark
stars were studied, the difference between the slow-rotation approximation and exact results 
for moderately rotating stars ($f\simeq 500$ Hz) at the canonical mass of 
$1.4\,\msun$ is of the order of 1 km. The discrepancy grows with 
stellar mass and spin frequency. 


\section{Transition to an exotic core: equilibrium, metastability, and instability}
\label{metastab}
\subsection{Thermodynamic equilibrium considerations}
\label{pure.mixed}
\begin{figure}
\resizebox{\hsize}{!}{\includegraphics*{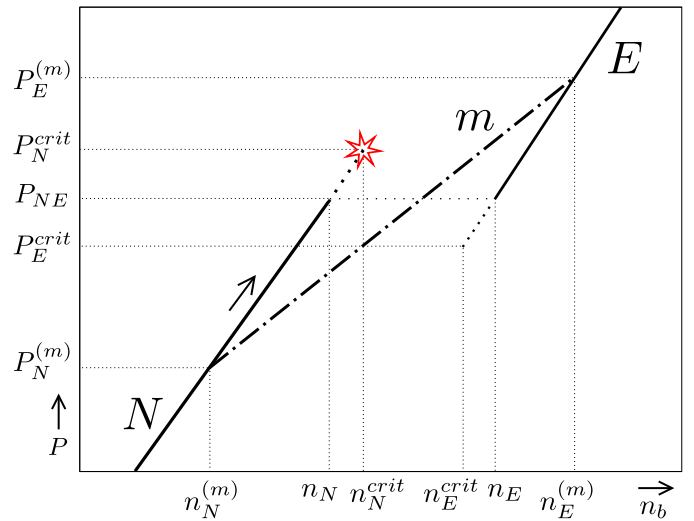}}
 \caption{Possible EOS  in the $n - P$ plane with various types of transition
between the N and E phases. \textit{ If the surface tension at the N-E
interface is large}: a mixed state is not present, and the N-phase is
stable up to $P_{\rm NE}$ (solid N-line). For $P>P_{\rm NE}$ and
$n>n_{\rm N}$ the N-phase is metastable with respect to the
nucleation of the E-phase (dotted line). The rate of nucleation
grows rapidly with overcompression $P-P_{\rm NE}$, and at $P=P_{\rm
N}^{\rm crit},n=n_{\rm N}^{\rm crit}$ the E-phase nucleates. 
After equilibration we get the E phase coexisting with the N phase at
pressure $P_{\rm NE}$, with the density jump $n_{\rm N}\longrightarrow
n_{\rm E}$ (thin dotted line horizontal segment, constant pressure).
After further compression the EOS continues in a pure E phase (thick
solid line).  \textit{ If the surface tension at the N-E interface is
small}: the equilibration  produces a mixed state of the E and N
phases, starting at $P_{\rm N}^{\rm (m)}$ (infinitesimal fraction of the
E phase) and ending at  $P_{\rm E}^{\rm (m)}$ with a vanishingly small
fraction of the N phase. The thick dot-dashed line $m$ is the mixed-phase segment
of the EOS. For $P>P_{\rm E}^{\rm (m)}$ we are dealing with a pure E
segment of the EOS. The continuous line N-m-E corresponds to the EOS
of matter in full thermodynamic equilibrium (see fig.~\ref{fig:mu-P}).
 The small dotted bottom segment of the E-line corresponds to the
metastable (with respect to the nucleation of the N-phase)
undercompressed E-phase.}
\label{fig:n-P}
\end{figure}

\begin{figure}
\resizebox{\hsize}{!}{\includegraphics*{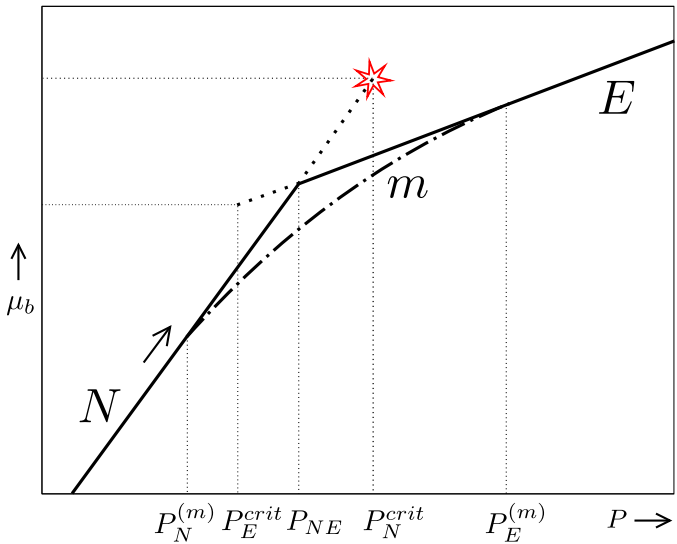}}
 \caption{Notation as in fig.~\ref{fig:n-P} but EOS in the $\mu_{\rm b}-P$
plane.}
\label{fig:mu-P}
\end{figure}

 Some theories predict that with increasing
density the NS core undergoes a transition from a normal (N) state to
a new exotic (E) phase. Some of these predicted phase transitions are
second order ones ({\it e.g.} kaon condensation, pion condensation), so that
density and composition of
 the matter are continuous at the transition point $P_{\rm t},n_{\rm t}$,
 while the speed of sound drops discontinuously $c_{\rm E}<c_{\rm N}$.
 However, for many models the softening in the  E phase just after threshold
 is so strong that it results in ${\rm d}P/{\rm d} n<0$, and therefore induces
 a density jump between the N and E phases, coexisting at some  $P_{\rm NE}$
 (see fig.~\ref{fig:n-P}). In this way, we have effectively  a first  order phase transition
 at constant $P=P_{\rm NE}$ between  the N phase  and the E phase. 
 This occurs for instance for a sufficiently
 strong pion or kaon condensation in nucleon matter. Another example of a  genuine 
 first  order phase transition is  quark deconfinement in dense hadronic
 matter. In general,
 in the first order phase transition at constant pressure, N and E phases are
 separated by a surface with a surface tension $\sigma>0$. We define the baryon chemical 
 potential in a given phase as $\mu_{\rm b} \equiv$ ${\rm d}{\cal E}/{\rm d}n_{\rm b}=(P+{\cal E})/n_{\rm b}$, 
 where ${\cal E}$ 
 is the energy density (including rest mass energy of particles). Thermodynamic equilibrium at 
 a given $P$ is realized by a state (phase) of the dense matter  with a minimum value 
 of $\mu_{\rm b}$.  However, it has been
 shown by Glendenning \cite{Glend92}, that if the surface contribution to the
 energy is sufficiently  small, the state of thermodynamic
 equilibrium ({\it i.e.} with minimum $\mu_{\rm b}(P)$)  is actually a mixture of coexisting phases E and N
 (mixed {\bf m} state)  in the pressure interval $P_{\rm N}^{\rm (m)}<P< P_{\rm E}^{\rm
 (m)}$. Such a mixed-phase state can be realized only for
 $\sigma<\sigma_{\rm max}$. Schematic plots of the EOS for the
 ${\rm N\longrightarrow \textbf{m}\longrightarrow E}$ and ${\rm N\longrightarrow
 E}$ realizations of the first order phase transitions to the exotic
 core are shown in figs.~\ref{fig:n-P} and \ref{fig:mu-P}.


\begin{figure}
\resizebox{\hsize}{!}{\includegraphics*{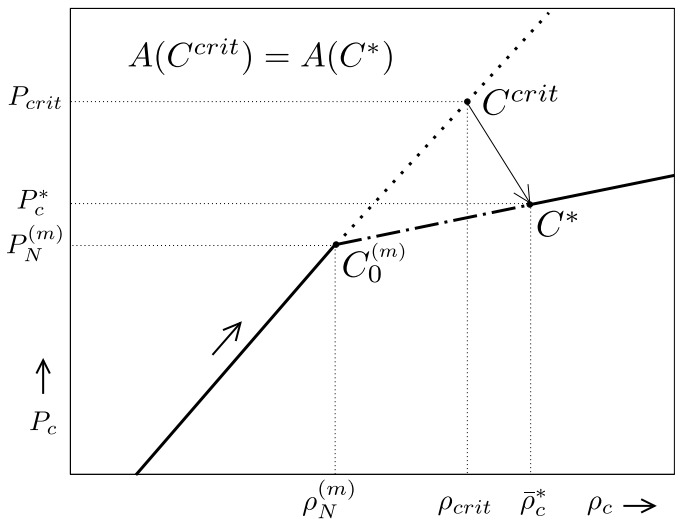}}
\caption{Trajectory of the neutron-star center in the $\rho_{\rm c} - P_{\rm
c}$ plane during spin-down or accretion, leading to a core-quake
after the nucleation of the E phase (configuration ${\cal C}^{\rm
crit}$) which implies a collapse into a configuration ${\cal C}^*$
with a mixed-phase core. The baryon number $A$ is conserved in
the collapse process. ${\cal C}_0$ is the last strictly stable
configuration with a N-phase core. For  further explanation see the
text.}
\label{fig:Pc-rhoc}
\end{figure}

\subsection{Central compression, metastability, and nucleation of the exotic phase}
\label{nucleation}
Consider  an element of matter at the center of a NS, consisting of the N-phase
at density $n_{\rm c}$ close to, but smaller than $n_{\rm N}^{\rm
(m)}$. The value of $n_{\rm c}$ can change  due to the NS evolution
induced by: (1) Angular momentum loss due to dipole radiation from
the radio pulsar; (2) Mass gain in the process
of matter accretion from a companion in a close binary system. In
most cases it is an increase of $n_{\rm c}$ (compression) that
proceeds  on a certain timescale $\tau_{\rm comp}=n_{\rm
c}/\dot{n}_{\rm c}$.  The increase of $n_{\rm c}$ 
during the spinning-down of an isolated pulsar depends on the 
mass of the star and reaches 5-30\% for evolution from the Keplerian to the 
non-rotating configuration \cite{Bejger11}. This increase however is proportional
to the square of $f_{\rm initial}/f_{\rm Kepler}$, and for an initial period
$\sim 10\;$ms it is less than   1\%. The timescales involved are longer than 
1 Gyr. In accreting NS the crucial parameter is the 
total angular momentum transferred to the star and the magnetic torque due to the 
interaction with an accretion disk (for details see \cite{Bejger11}). The central
compression is small (few percent) in the absence of a magnetic torque and 
for accretion from the marginally stable orbit, but could be as large as 10\%
for $B\sim 5\times10^8$~G after accretion of $0.1\;\msun$. It would take 
$10\;$Myr for a NS accreting at the rate $10^{-8}\;\msun/{\rm yr}$ in a low mass
X-ray binary. The compression timescale becomes rather short (years) for a very 
specific class of young magnetars  (see  \cite{Dimmel2009} and references 
therein). 

In scenarios (1)-(2) the temperature at the NS center is $<10^{9}~$K
and therefore thermal contributions to thermodynamic quantities are
small: matter is strongly degenerate, and thermal fluctuations are
negligibly small compared to the energy barriers separating the N
and {\bf m} states. In any case, the transition to the {\bf m} state has to be
initiated by a droplet of the E phase. Even neglecting the  surface tension
contribution to the energy of an E-drop ($\sigma=0$), it is
energetically possible only for $n_{\rm b}>n_{\rm N}$ (at lower $n_{\rm b}$ the drop
decays back into the N-phase). However, to nucleate the E-phase in the
N-medium, an energy barrier, created actually by the surface
contribution, has to be quantum-mechanically penetrated with the energy
supply coming from quantum fluctuations. After an  E-drop nucleates,
it grows into a bulk E-phase which coexists stably with the N-phase at
$P_{\rm NE}$, with a density jump $n_{\rm N}\longrightarrow n_{\rm E}$
at the phase interface.

Sometimes it may be convenient to visualize the evolution of the NS
center in the $P - \mu_{\rm b}$ plane, fig.~\ref{fig:mu-P}. Assume
that the central core is being compressed, so that $P_{\rm c}$ grows
in one of the astrophysical processes described in the first paragraph
of the present section. Compression corresponds to a trajectory in
the $P - \mu_{\rm b}$ plane. Even after passing $P_{\rm c}=P_{\rm
N}^{\rm (m)}$, the core with $P_{\rm c}>P>P_{\rm N}^{\rm (m)}$
consists of a pure N-phase and grows in time, because the mixed state
\textbf{m} cannot be reached due to the impossibility of nucleation
of the E-phase because of the energy barriers (resulting from
surface tension and Coulomb interaction). After reaching $P_{\rm
c}=P_{\rm NE}$, the star's center enters the metastable
(overcompressed) segment $P>P_{\rm NE}$ of the N-phase EOS. The
lifetime with respect to nucleation of the E-phase $\tau_{\rm nucl}$
decreases rapidly with growing overpressure $\Delta P = P_{\rm c} -
P_{\rm NE}$. Nucleation of quark matter in dense baryon cores  
was studied in \cite{Berezhiani2003,IidaSato1997,IidaSato1998}, while 
the nucleation of the pion-condensed state was discussed in 
\cite{HaenselSchaeffer1982,Tatsumi1987,Muto1990}. As soon as 
$\tau_{\rm nucl}\sim\tau_{\rm comp}$ (which takes place at 
$P_{\rm c}=P_{\rm crit}$), droplets of
the E-phase appear spontaneously and grow into  regions of the
E-phase coexisting with the N-phase. This kinetic (non-equilibrium)
process implies a local pressure deficit and a collapse of the central
core. If the heat release is sufficiently large, one
can contemplate a redistribution and coagulation of the E-droplets  in
the N-phase, creating a \textbf{m}-state core, larger than a uniform
E-core could have been.

Such a situation is schematically depicted in fig.~\ref{fig:Pc-rhoc}.
The central core is being compressed while still in the N-phase,
until at $P_{\rm c}=P_{\rm crit}$ the E-phase  nucleates in the
N-medium. Assuming that thermodynamic fluctuations are sufficiently
strong, one gets a mixed \textbf{m} state extending down to the
$\left(P_{\rm N}^{\rm (m)},\rho_{\rm N}^{\rm (m)}\right)$ point.
This corresponds to a full thermodynamic equilibrium in the core.
However, before this final state has been reached, nucleation of the
E-phase at $P_{\rm crit}$ implied a local pressure deficit, and a
collapse of a NS into a new more compact configuration took place.
After the E and N phases mix, a large \textbf{m}
core is formed in configuration ${\cal C}^*$, with a mean central
density $\overline{\rho}^*_{\rm c}>\rho_{\rm crit}$ and $P^*_{\rm
c}<P_{\rm crit}$.
 The dynamics of this minicollapse process induced by a core-quake 
is discussed in sect. \ref{sect:minicoll}.

\section{Exotic cores and NS parameters}
\label{exoNS}

\subsection{Hyperonic cores}
\label{sect:hypNS}
Hyperons (baryons containing at least one strange quark) were discovered in
laboratory in the early 1950s and are studied experimentally since then. In the
late 1950s it was suggested that hyperons could also be present in NS cores
(see \cite{HPY2007book,Isaac} for a historical perspective). Indeed, although
hyperons are unstable under terrestrial conditions, at densities typical for 
the NS centers, the Pauli exclusion principle prevents them from decaying
into nucleons. 

Relatively little is known about the properties of the interactions of hyperons
with other baryons from hyperon scattering (see discussion in \cite{S08,OP15}).
On the one hand, thanks to the study of $\Lambda$-hypernuclei and
$\Xi$-hypernuclear states in laboratory, the potential for the $\Lambda$ and
$\Xi$ hyperons in symmetric nuclear matter at saturation density is found to be
attractive. But on the other hand, contradicting results were found for the
$\Sigma$ potential. Moreover, only few double-$\Lambda$ hypernuclei were
studied indicating an attractive $\Lambda - \Lambda$ potential and no other
pairs of hyperons as $\Lambda-\Xi$ or $\Xi-\Xi$ were observed.

\begin{figure}
\resizebox{\hsize}{!}{\includegraphics*{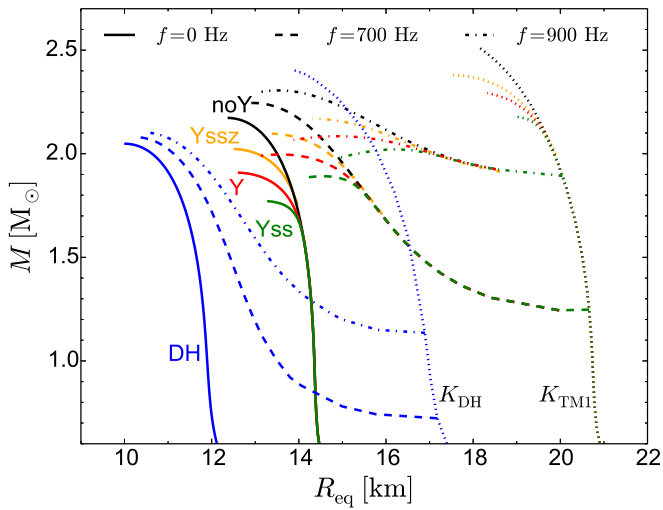}}
\caption{(Colour online) $M-R_{\rm eq}$ relations for non-rotating and rotating stars, at 700 and 900 Hz. Results are shown for the DH (unified) EOS  and various versions of the TM1 one, with three different composition in the inner core. See text for details.}
\label{fig:TM1content}
\end{figure}

Allowing for a possible transition to hyperonic matter results in a
softening of the EOS at densities $n_{\rm b} > 2n_0$ and, as a
consequence in a decrease of the maximum mass. This is illustrated in fig.~\ref{fig:TM1content}, where
$M-R$ relations at various spin frequencies are plotted for the DH EOS and for
the EOS obtained within the RMF (relativistic mean field) approach with the TM1 model \cite{TM1}, as an example. For the
latter four different EOS for the core are shown. The first model {\bf noY}
corresponds to a purely nucleonic composition. Three models allowing for a 
transition  to hyperonic matter at high density are plotted. In the {\bf Y} model,
 the vector-isoscalar hidden-strangeness $\phi$
mesons is included and vector-mesons couplings to baryons are given by the SU(6) symmetry
following \cite{MC13}. The $\phi$ meson which is coupled to hyperons only, 
 yields an  additional repulsion between hyperons and thus leads to a
stiffening of the EOS and an increase of the maximum mass. The {\bf Yss} includes
 in addition to the $\phi$ meson, the scalar-isoscalar hidden-strangeness $\sigma^*$ meson
in SU(6) symmetry following \cite{MC13}. This meson which also couples to
hyperons only, enables reproducing  a  weakly attractive $\Lambda - \Lambda$
potential \cite{SD94}. The consequence of its inclusion is a mild softening of
the EOS and thus a slight decrease of the maximum mass as can be seen from
fig.~\ref{fig:TM1content} for non-rotating stars. 
Finally, the {\bf Yssz} model corresponds to the {\bf Yss} one except that SU(6) 
symmetry is broken following \cite{GusakovHK2014}. The effect of this breaking is 
studied in detail in {\it e.g.} \cite{Weis12}. The specific choice of parameters in the {\bf Yssz} model,
 results in a stiffening of the EOS, which becomes even stiffer 
than the {\bf Y}  model although $\sigma^*$ mesons are included. As a consequence the maximum mass 
is the highest of all the hyperonic models but nevertheless lower than the one for a purely nucleonic star.

 Observations of massive neutron stars with a mass $2\;\msun$ \cite{Demorest2010,Antoniadis2013} are 
therefore challenging for hyperonic EOS. Reconciling the possibility of a
transition to hyperonic matter at high density with observations of massive NS
requires solving  the so-called `hyperon puzzle'. In \cite{FortinZHB2015Rhyp},
a systematic study of all EOS for hyperonic matter, consistent
with a $2\;\msun$ maximum mass,  available at the time of publication, was
conducted (14 EOS, all but one being RMF models). It was
shown that all of them give  pressures in pure neutron matter at densities close to
$n_0$  which are too large compared with recent precise many-body calculations for pure
neutron matter. These calculations were performed using two different approaches:
 quantum Monte Carlo method \cite{Gandolfi2012} and
chiral effective field theory \cite{Hebeler2013}, and are 
in a remarkable mutual agreement. A large pressure for $n_{\rm b}\lesssim n_0 \lesssim 2n_0$ 
in fact  is needed to balance  the hyperon softening at higher density, and is
correlated with large radii: $R>13$~km for neutron stars with masses
$M=1.0-1.6\;\msun$. Hyperonic EOS consistent with a $2\;\msun$ maximum mass 
and with a pressure close to $n_0$ consistent with \cite{Gandolfi2012,Hebeler2013}
 are obtained in {\it e.g.} \cite{OP15,PFGR}.
 In \cite{KV05} a RMF model, so-called KVOR, where hadron masses and coupling constants are scaled by functions depending on the
scalar field is formulated.
Hyperonic NS with $M\geq 2\;\msun$ are obtained in two versions of this model. The MKVORH one \cite{Maslov2015hypa} 
assumes a relatively low value of the nucleon effective mass at saturation and yields $M_{\rm
max}=2.2-2.3\;\msun$ and $R_{1.4}=12.2\;$~km while the KVORcut3 version \cite{Maslov2015hypb} 
has higher effective mass at $n_0$ and $M_{\rm max}=2.0-2.3\;\msun$ for $R_{1.4}=13.0\;$~km. However, even at maximum mass
the strangeness per baryon is very small: $\lesssim 3\times10^{-2}$. 
A solution to the ``hyperon puzzle'' is therefore
reached in \cite{Maslov2015hypa,Maslov2015hypb} when there are nearly no hyperons.

Fig.~\ref{fig:TM1content} 
also presents $M-R$ relations for NS rotating at $700$ and $900$~Hz together
with $M-R$ relation for NS rotating at the mass-shedding limit $f_{\rm K}$. Let
us now compare results for the DH and TM1 EOS. On the one hand, the increase of
the maximum mass due to rotation is larger for the TM1 EOS, with larger radii
for non-rotating configurations, than for the DH EOS. On the other hand, the
increase of the minimum mass, which is located at the intersection between the
$f_{\rm K}$ curve and the $M-R$ one, is larger for the TM1 EOS, for a given
rotation rate. In other words the DH EOS, which has smaller radii for
non-rotating models than the TM1 one, has for a given rotation rate a broader
range of masses than TM1. Under the effect of rotation the $M-R$ relation for
TM1 become flatter than the DH EOS. The property that 
$M-R$ relations become flatter for EOS with larger radii for non-rotating configurations 
than for those with smaller
radii is even more dramatic for very high spin frequency, close to the
Keplerian frequency \cite{BejgerHZ2007}. For the TM1 model, the softer the high
density part ({\it e.g.} when comparing the {\bf Yss} EOS to the {\bf Y} one)
, the smaller the increase of $M_{\rm max}$ and the
larger the increase of $R(M_{\rm max})$ with rotation. As a consequence for a
given rotation rate, the $M-R$ curve is flatter for the softer TM1 EOS and the
narrower is the stellar mass range.

\begin{figure}
\resizebox{\hsize}{!}{\includegraphics*{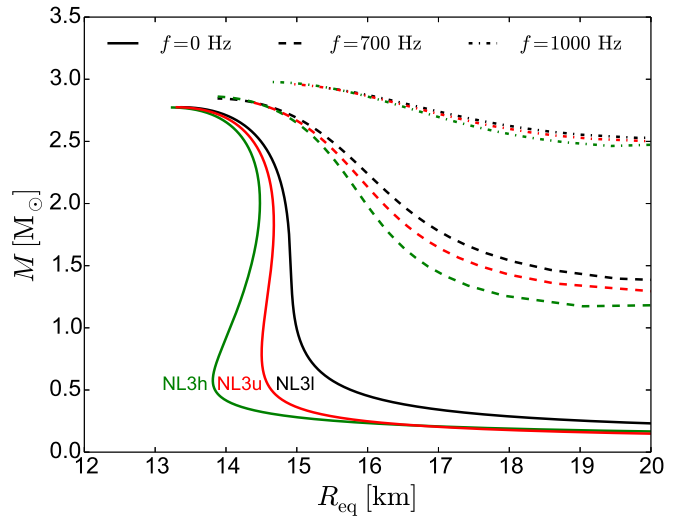}}
\caption{(Colour online) $M-R_{\rm eq}$ relations for non-rotating and rotating stars, at 700 and 900 Hz. The NL3 EOS is used for the core. Three different core-crust matchings are used: one using a unified crust (NL3u) and two using the DH EOS for the crust (NL3l and NL3h). See text for details.}
\label{fig:NL3matching}
\end{figure}

There exists very few unified EOS, in the sense that the same nuclear
interaction model is used to describe both the clusterized matter in the crust
and the homogeneous one in the core \cite{DouchinH2001,FC13}. Two non-unified
EOS for the crust and the core are usually `glued' together by ensuring that
the pressure $P(n_{\rm b})$ and energy density $\rho(n_{\rm b})$ are increasing
functions of the baryon number density $n_{\rm b}$. However there is no unique
prescription for the transition between two different EOS \cite{PFGR}.  For
example in fig.~\ref{fig:NL3matching} three possible choices of
`gluing' for the same crust and core EOS: the DH \cite{DouchinH2001} and NL3
RMF models \cite{NL3} are shown. The NL3l EOS corresponds to gluing DH and
NL3 at the density at which the $P(n_{\rm b})$ relations for the two EOS cross:
$n_{\rm b}=0.046$ fm$^{-3}$. For the NL3h EOS, the core EOS is glued to the
crust at $n_{\rm b}=0.16$ fm$^{-3}$. Finally, the EOS NL3u corresponds to using
the same nuclear model for both the crust and the core for the NL3
parametrization. The crust model is taken from \cite{GP}, where the Thomas-Fermi
approach is used to describe the clusterized matter in the crust. Similarly, 
unified EOS for the TM1 model are shown in fig.~\ref{fig:TM1content}. 
Fig.~\ref{fig:NL3matching} shows the $M-R$ relations obtained for the
three NL3 EOS: for non-rotating stars the difference in the equatorial radii
between the EOS decreases when the mass increases but can be as large as $4\%$
of the radius for a $1.4\;\msun$ NS and $3\%$ for a $1.6\;\msun$ NS. For a
given mass the difference in radii between various matching prescriptions
increases with spin frequency. For example for a NS rotating at $700$~Hz the
difference in radii increases to $13\%$ and $6\%$ of the radii for a $1.4$ and
$1.6\;\msun$ NS respectively. Therefore calculations of 
unified EOS is of great importance in order to properly describe the macrophysical properties of NS \cite{PFGR}.

\subsection{Hybrid stars}

Some theories of dense hadronic matter predict a {\it deconfinement} of
quarks  at densities achievable in the cores of massive NS. 
The phase transition from the baryon phase of matter (N)
to the deconfined quark matter (Q) is usually assumed to be of first
order. It softens the EOS due to the density jump at constant
$P_{\rm NQ}$ (transition between pure N and Q phases with a density jump 
at the interface $\lambda = \rho_{\rm Q}/\rho_{\rm N}$), or via a
mixed-phase region. In spite of this softening, the existence of
$2\;\msun$ pulsars does not exclude quark cores in NS, but imposes
rather tight constraints on the EOS with N-Q transition
\cite{EMMI2014,Klaehn2013}. 
Two subsequent phase transitions through the intermediate quark phase 
energetically preferred in a rather narrow range of densities were also considered \cite{Pagliara08,Blaschke10,Zdunik13,AC13}. 

In all these cases the density jump (or two jumps) is the main source of softening of matter at the transition pressure $\pne$ determining $M(P_{\rm c})$, the $M(R)$ dependence in the vicinity of the configuration
${\cal C}_0$ with central pressure equal to the phase transition pressure $P_{\rm c}=\pne$.

It should be mentioned that the baryon-quark phase transition between two phases can proceed through a mixed phase
\cite{Glen2000,Glend92} in which the condition of local charge neutrality (for the two phases separately) is relaxed.
Although the properties of stars (mass-radius relation) are not the same for an EOS with phase transition involving a mixed or pure phases, 
significant differences are observed for a rather small region of central pressures (close to the transition pressure).
Global properties such as the existence of instability regions, ``twins'' and maximum mass of hybrid star are
very similar \cite{Alford05}.
 
The conditions for the quark EOS resulting from the maximum mass constraint ($M_{\rm max}>2\msun$) are the following:
the softening effect of the first order phase transition has to be compensated by a stiff quark EOS (larger $\lambda$
corresponds to  a larger sound velocity $c_{\rm Q}$ of the quark phase, \cite{Zdunik13}).  
The EOS for the baryon phase should also be relatively stiff - 
the configuration ${\cal C}_0$ cannot be too compact and close to the maximum mass for 
the N-phase stars. If the latter 
requirement is not fulfilled, the phase transition at the center leads almost immediately to the dynamical instability and
gravitational collapse of a hybrid star. For phase transitions to quark matter at densities $2.5 - 3.5 n_0$, the 
sound velocity in the quark phase\footnote{These results were obtained for a simplified quark EOS with constant sound velocity. It was shown however, that this form of $P(\rho)$ dependence approximates well the 
EOSs obtained for more sophisticated, microscopic models of quark matter \cite{Zdunik00,Zdunik13}.} should be larger than $0.6c$. 

Benic {\it et al.} \cite{Benic2015twin} presented the possibility of the existence of high-mass twins - two families of dense objects (baryon
and hybrid stars) with masses about $2\;\msun$. In their model two conditions discussed before are fulfilled - the matter
in the NJL8 quark phase is stiff ($c_s\sim 0.6 - 0.9 c$), and the compactness of the ${\cal C}_0$ configuration for $\simeq 2\;\msun$ is 
not large, but comparable with the compactness of a typical nucleon NS with $M=1.4\;\msun$ and $R\simeq 11\;{\rm km}$. 

\begin{figure}
\resizebox{\hsize}{!}{\includegraphics*{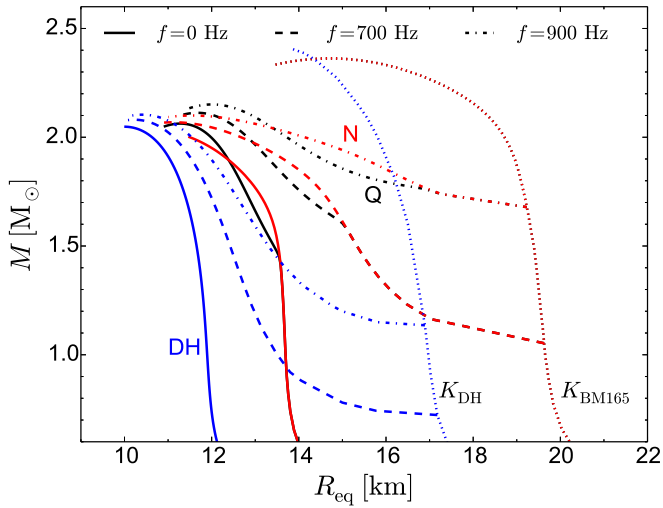}}
\caption{(Colour online) DH (unified) EOS and BM165 \cite{BednarekHZBM2012} with and without phase transition to quark matter. Quark phase is
approximated by linear dependence $P=a\cdot(\rho-b)$ with $a=0.5c^2$ and the density jump at the phase transition 
$\lambda=1.2$. For details see \cite{Zdunik00,Zdunik13}.}
\label{fig:BM165Q}
\end{figure}

In fig.~\ref{fig:BM165Q} we present the $M-R_{\rm eq}$ relations for the model with a first order phase transition
to quark matter (Q, the black curve). The effect of rotation for hybrid stars is similar to that of hyperon stars
(fig.~\ref{fig:TM1content}), although in this case the relatively large stiffness of the quark matter EOS leads to the maximum
mass larger than for a baryon star. The minimum mass of the  hybrid star rotating with $f=700$~Hz is larger than $1\;\msun$, much
larger than for the DH EOS, and comparable to the hyperon stars discussed in sect.~\ref{sect:hypNS}. We should stress however
that this minimum mass (Keplerian configuration for $f=700$~Hz) corresponds to the star composed entirely of non-strange,
nucleon matter  ($P_{\rm c}<\pne$). The large value of $M_{\rm min}$ is an indirect 
consequence of the EOS softening at high density and the $M_{\rm max}>2\;\msun$ requirement, which results in a large radius
({\it i.e.,} small compactness) of the ${\cal C}_0$ configuration.

\section{Limits on rigid rotation}
\label{kepler}
\subsection{Keplerian (mass-shedding) limit}
\label{mass-shedding}
\begin{figure}
\resizebox{\hsize}{!}{\includegraphics*{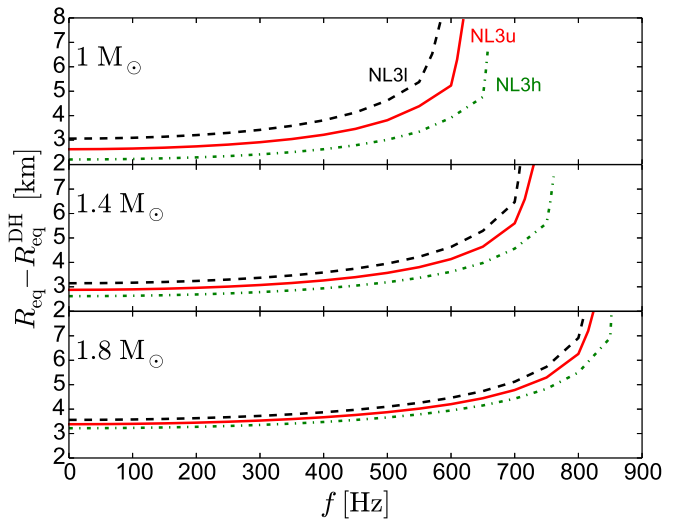}}
\caption{(Color online) For a given mass, as a function of the spin frequency $f$, difference in $R_{\rm eq}$ between configurations for the DH EOS and different matchings for the NL3 EOS. See fig.~9 and section 5.1 for details.}
\label{fig:NL3-DH}
\end{figure}

Consider a static ($f=0$) NS of gravitational mass $M_{\rm s}$ and
baryon mass $M_{\rm b}$. Then construct a sequence of rigidly
rotating configurations with the same $M_{\rm b}$ and increasing
$f$. This sequence will terminate at the mass-shedding limit with 
the Keplerian frequency $f_{\rm K}$. At this limit the spin frequency of the NS
is equal to the orbital frequency of a test particle on a circular orbit 
corresponding to the NS equator. For $f>f_{\rm K}$
hydrostatic equilibria of a NS with baryon mass $M_{\rm b}$ do not
exist.

As we have seen in sect.~\ref{exoNS}, the EOS that predicts
hyperonization  and therefore a softening after the hyperon threshold at
 $\rho\simeq 2\rho_0$, has to contain  a sufficiently stiff nucleon segment
$\rho_0<\rho<2\rho_0$,  in order to get $M_{\rm max}>2~\msun$ in
spite of the hyperon softening.  This implies that the radii of the
pre-hyperonic NS  with $\rho_0<\rho_{\rm c}<2\rho_0$, and in
particular the radius of $1.4\;\msun$ NS  with this EOS, $R^{\rm (H)}_{1.4}$, 
are expected to be larger
than those of ``standard'' NS with nucleon cores, $R_{1.4}^{\rm (N)}$. 
A rough Newtonian argument about the centrifugal
force $\propto R_{\rm eq}^2$, predicts therefore that for the 
same moderate $M_{\rm s}$ mass: 1) the Keplerian frequency for EOS 
with hyperonization  is smaller than for 
purely nucleonic EOS, 2) the difference  $R^{\rm(H)}_{\rm eq}-
R^{\rm(N)}_{\rm eq}$ grows with increasing $f$ (see fig.~\ref{fig:NL3-DH}). 
It should be 
stressed that both properties are
valid  for $M<0.9\;M_{\rm max}^{\rm (stat)}$ \cite{FortinZHB2015Rhyp}.

The results reviewed above can be corroborated quantitatively by
theoretical considerations. For a non-rotating spherically symmetric 
NS of gravitational mass $M$ and circumferential radius
$R$ the orbital frequency of a test particle in a circular orbit of
radius $r_{\rm orb}>R$ is \cite{ST1983}
\begin{equation}
f_{\rm orb}={1\over 2\pi}\left({GM\over r^3_{\rm
orb}}\right)^{1/2}~. \label{eq:f.orb}
\end{equation}
According to the Birkhoff theorem the formula is the same as for a central
point mass $M$. For a static, spherically-symmetric NS of mass $M$
we therefore get
\begin{equation}
f_{\rm orb}=f^{\rm sph}_{\rm K}={1\over 2\pi}\left({GM\over
R^3}\right)^{1/2}~. \label{eq:fK.sph}
\end{equation}
which is identical with the Newtonian formula for a mass-shedding
limit for a spherically symmetric  self-gravitating star.

It has been shown, with a precision of a few percent, that the
formula for $f_{\rm K}$, eq.~(\ref{eq:fK.sph}) holds also for
realistic NS at the Keplerian limit provided we replace $R$ by the
equatorial radius at the Keplerian limit $R_{\rm K}$ 
\cite{BejgerHZ2007}. The high precision comes as a surprise, because a NS 
at the Keplerian limit is strongly flattened and  rapid rotation produces
strong frame-dragging effects in the exterior space-time. The formula
\begin{equation}
f_{\rm K}={1\over 2\pi}\left({GM\over R_{\rm K}^3}\right)^{1/2}~, 
\label{eq:fK.magic}
\end{equation}
which is (surprisingly) so precise for NS, but not close to maximum
allowable mass:  $M_{\rm s}<0.9\;M_{\rm max}^{\rm stat}$, holds
strictly for the relativistic Roche model with extreme central condensation of
mass \cite{Shapiro1983,Shapiro1989,HZBLattimer2009}. It
should be stressed that validity of this model breaks down near 
$M^{\rm stat}_{\rm max}$. Moreover, the formula (\ref{eq:fK.magic})
is much less precise for strange quark stars built of self-bound quark matter which are
characterized by a rather uniform density \cite{HZBLattimer2009}.

 For practical applications, one can use the 
empirical formula of ref.~\cite{HZBLattimer2009}, valid for NS with
and without an exotic core:

\begin{equation}
f_{\rm K}(M_{\rm s})\approx 1.08~{\rm kHz}\;\left({M_{\rm s}\over
\msun}\right)^{1/2}\left( R_{\rm s}\over 10~{\rm km}
\right)^{-3/2}, \label{eq:fK.empir}
\end{equation}
where $M_{\rm s}$ and $R_{\rm s}$ are mass and radius of a static
configuration of the same baryon mass as the Keplerian
configuration. It holds for  $0.5~\msun<M_{\rm s}<0.9 M_{\rm
max}^{\rm stat}$ and implies $f^{\rm (H)}_{\rm K}(M_{\rm s})<f^{\rm
(N)}_{\rm K}(M_{\rm s})$ for $M<1.6\;\msun$. Note that the
empirical prefactor $1.08~{\rm kHz}$ is larger
than $1.00~{\rm kHz}$ which corresponds to the relativistic Roche model.

\subsection{Maximum frequency of stable rotation}
\label{fmax.stab}
The empirical formula for the absolute maximum  of $f$  for stably 
rotating configurations with a given EOS, $f_{\rm
max}$ (\cite{HPY2007book} and references therein), is of a
different character than eq.~(\ref{eq:fK.empir}). 
Namely, the formula for $f_{\rm max}$
results from an approximate but quite precise correspondence between
two extremal configurations:  the static configuration with a
maximum allowable mass, $M_{\rm max}^{\rm stat}, R_{M_{\rm max}^{\rm
stat}}$, and another extremal configuration, which is stably
rotating with a maximum allowed frequency (called maximally-rotating
configuration). This extremal configuration is  stable  with respect to 
mass-shedding and stable with respect to axisymmetric perturbations 
(large filled dots in fig.~\ref{fig:mr-static-kepler}). 

\begin{equation}
f_{\rm max}[{\rm EOS}]\approx 1.22~{\rm kHz}\;\left({M_{\rm max}^{\rm
stat}\over \msun}\right)^{1/2}\left({R_{M_{\rm max}^{\rm stat}}
\over 10~{\rm km}} \right)^{-3/2}~. 
\label{eq:fmax.empir}
\end{equation}
The prefactor $1.22~{\rm kHz}$ is significantly larger than
$1.08~{\rm kHz}$ in eq.~(\ref{eq:fK.empir}). Moreover, in contrast to
eq.~(\ref{eq:fK.empir}), the formula for $f_{\rm max}$ is valid for
both NS and self-bound quark matter stars. However, it can be used 
to constrain only static
configuration with the maximum allowable mass. Assuming $M_{\rm
max}^{\rm stat}=2.0\;\msun$ and $R_{M_{\rm max}^{\rm stat}}=10~{\rm
km}$ one gets $f_{\rm max}=1725~{\rm Hz}$.

\begin{figure}
\resizebox{\hsize}{!}{\includegraphics*{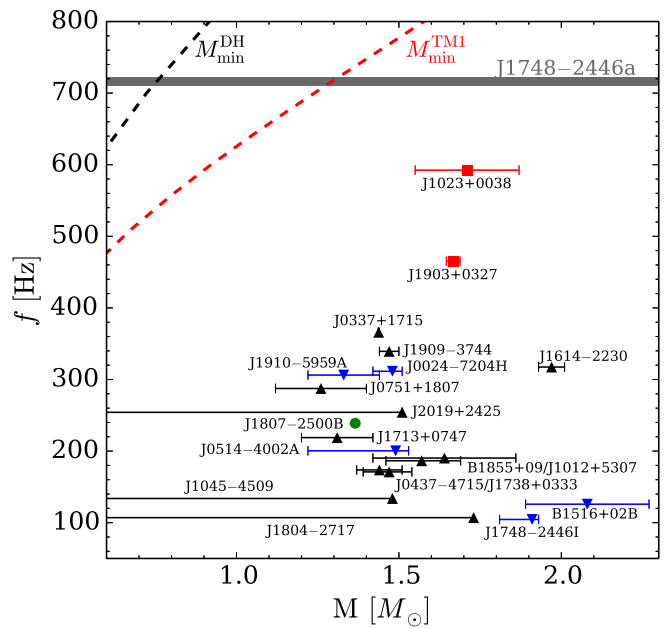}}
\caption{(Color online) Observed frequency $f$ {\it vs.} measured mass $M$ of radio MSP (with $f>100$~Hz). Observational data: see Tab.~\ref{tab:mf} in the Appendix. Black up-pointing triangles: NS in Galactic binaries with a white dwarf (WD); blue down-pointing triangles: NS in a binary with a WD in a globular cluster (GC); red squares: NS in Galactic binaries with a main-sequence star; green dot: NS in a binary with a NS in a GC. No NS with measured mass  in a Galactic  NS-NS binary rotates with $f>100$~Hz. The grey line indicates the measured frequency of the fastest rotating  NS, PSR J1748$-$2446a. The dashed lines correspond to the relation between the minimum mass of rotating NS and the frequency for the DH EOS (in black) and the TM1 EOS (in red).}
\label{fig:Mvsf}
\end{figure}

\subsection{Constraint from $f_{\rm max}^{\rm obs}=716\;$Hz}
\label{mass716Hz}

Figs.~\ref{fig:mr-static-kepler}, \ref{fig:TM1content} and \ref{fig:BM165Q} readily 
illustrate that the minimum mass of a rotating NS, $M_{\rm min}^f$, is sensitive both to $f$ 
and to the EOS. This is to be contrasted with the minimum mass of a static NS, which is
rather weakly dependent on the EOS (provided it is a unified EOS, so that the crust 
and liquid core are calculated using the same nuclear interaction model). We get 
$M_{\rm min}^{\rm stat}\simeq 0.1\;\msun$ (see \cite{HPY2007book} and references  
therein). 

 In fig.~\ref{fig:Mvsf} the relations $M_{\rm min}(f)$ obtained for two unified EOS: TM1 and DH are plotted against the frequency and mass\footnote{See {\it e.g.} {\tt http://stellarcollapse.org/nsmasses}, {\tt www3.mpifr-bonn.mpg.de/staff/pfreire/NS\_masses.html} or \cite{ZW11}.} of observed NS with $f>100$~Hz, together with the maximum observed frequency. This figure, figs.~\ref{fig:TM1content} and \ref{fig:BM165Q}
show how sensitive $M_{_{\rm min}}^{^{\rm 716\;Hz}} $ is to the nucleonic segment of the EOS for $n_{\rm b}\lesssim 
2n_0$. For the DH EOS, $M>M_{_{\rm min}}^{^{\rm 716\;Hz}} =0.75\;\msun$ while for the stiff TM1 EOS: $M>M_{_{\rm min}}^{^{\rm 716\;Hz}} =1.29\;\msun$. As a consequence, the observation of a fast rotating NS with a low mass could potentially constrain the EOS. For example if a NS is observed with $M< 1.5\;\msun$ and $f=800$~Hz, then the TM1 EOS would be ruled out. Nevertheless, for the time being, no current observation provide such a constraint, as shown in fig~\ref{fig:Mvsf}.

\section{Rotation, stability and dynamics}
\label{dynamics}

\subsection{Back-bending and stability of rotating stars}
\label{sect:bbstab}

\begin{figure}

\resizebox{\hsize}{!}{\includegraphics{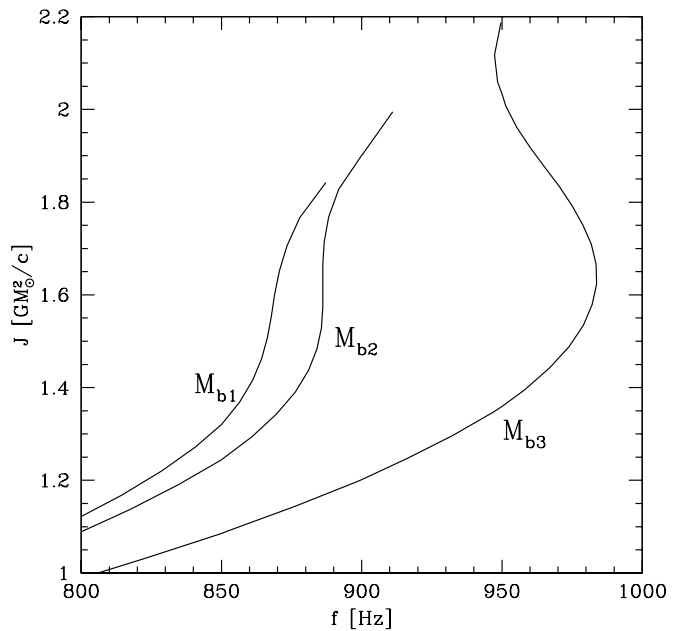}}
\caption{The back-bending phenomenon for an EOS softened by a mixed phase at high density. Shown are evolutionary tracks 
for an isolated NS loosing its energy and angular momentum $J$ (the total baryon mass is fixed along each curve,
$M_{\rm b1}<M_{\rm b2}<M_{\rm b3}$). For a sufficiently large mass ($M>M_{\rm b2}$), spin-up due to angular momentum loss is observed.}
\label{fig:jfx}
\end{figure}

The appearance of a new phase at the center of a NS results always in 
a softening of the EOS. 
For the global properties of NS the consequence of this softening is
a slower increase of the stellar mass and total baryon mass as central pressure
increases.
For significantly strong softening of the EOS  and sufficiently large star mass this feature can be observed in the case of
the evolution of an isolated pulsar as the so-called back-bending phenomenon - the epoch
at which the angular momentum loss due to the evolutionary processes leads to
the spin up of the star. 
As a rigidly-rotating star of a fixed total baryon mass $M_{\rm b}$ looses its angular momentum $J$, the central pressure and density increase.
For a strong softening of the equation of state above a critical density it is possible (for some range of 
$M_{\rm b}$) that 
for a slowing-down star the central density crosses this critical value and 
the core of a new, dense phase develops in the center. Then the  star shrinks with a significant decrease of the moment of inertia $I$ and this  has to
be compensated by the increase of rotational frequency $\Omega$ to fulfill the equality $\delta J=\Omega\delta I+ I\delta\Omega$.
 The back-bending phenomenon (associated  in fig.~\ref{fig:jfx}
with an ``S'' shape of the $J(f)$ curve) was proposed 
in \cite{Glen97} as a signature of a phase transition to an  exotic state 
of matter in the center of a spinning-down pulsar (for example the appearance of a
mixed-phase  core at the star center). It should be however mentioned that a similar behavior
can be also caused by hyperonization of dense matter, 
provided that the EOS softening above the  hyperon threshold is strong enough \cite{Zdunik04}. 

\begin{figure}

\resizebox{\hsize}{!}{%
  \includegraphics{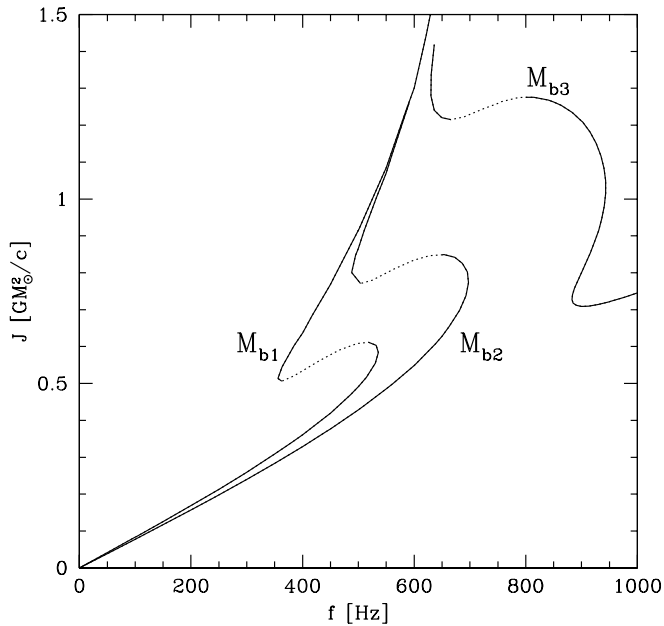}
}
\caption{Total stellar angular momentum $J$ as a function of rotational frequency $f$ for evolutionary tracks of an isolated NS with fixed baryon mass. The softening of the EOS is stronger than in the case presented in fig.~\ref{fig:jfx}, which results in unstable region where $J$ increases with increasing central density (dotted line).}
\label{fig:jfxi}
\end{figure}

The back-bending phenomenon is not the only consequence of the softening of the dense matter EOS. For a significant softening the result could be even more spectacular - a region of
configurations which are dynamically unstable appears. 
Increasing the softening of the EOS finally results in a non-monotonic behavior of $J$ along
an evolutionary track with a fixed $\Mb$. Configurations in a  region where $J$ increases with increasing $\rho_c$ are  unstable with respect to small axisymmetric 
perturbations (fig.~\ref{fig:jfxi}).

The discussion of the existence of the back-bending phenomenon or the stability of the rigidly rotating
configurations can be performed by the analysis of the extrema of three basic,
macroscopic parameters of the rotating star that are well defined in GR - $M$, $\Mb$ and $J$ as functions of 
any variable which parametrizes stationary rotating stellar configurations,  for example
central density $\rho_{\rm c}$, pressure $P_{\rm c}$ or equatorial radius $R_{\rm eq}$.

 The change in stability  of a rigidly rotating configuration 
(from stable to unstable or {\it vice versa}) 
corresponds to an extremum of the two parameters from the $\{ M,\Mb, J\}$ set, 
with the third parameter fixed \cite{FriedmanIS1988}. 
 According to the turning-point theorem, the stability of a rotating configuration 
can be stated by checking that
\begin{equation}
\left({\partial M\over \partial \pc}\right)_{J}>0,\quad
\left({\partial \Mb \over \partial \pc}\right)_{J} >0,\quad
\left({\partial J\over \partial\pc}\right)_{\Mb} <0. 
\label{eq:M.MB.J.stab}
\end{equation}
The back-bending phenomenon corresponds to the existence of a region  
in which, due to some evolutionary processes, the rotational frequency $f$ 
increases with a decreasing total angular momentum $J$. It can be written as 
\begin{equation}
\left({\partial J\over \partial f}\right)_{\Mb} < 0, 
\label{eq:J.MB.bb}
\end{equation}
or equivalently
\begin{equation}
\left({\partial \Mb\over \partial \pc}\right)_{f} < 0. 
\label{eq:MB.f.bb}
\end{equation}

The condition for the  back-bending phenomenon in
eq.~(\ref{eq:MB.f.bb}) is similar to the second relation
in eq.~(\ref{eq:M.MB.J.stab}) with $J$ replaced by $f$. 
The back-bending manifests 
itself in the existence of a local minimum of the $\Mb(x)$ curves plotted
for fixed frequency $f$. However in order to decide if the configurations which 
are subject to back-bending are stable one has to analyze $\Mb(x)$ relations 
for fixed total angular momentum $J$. 

\begin{figure}

\resizebox{\hsize}{!}{%
  \includegraphics{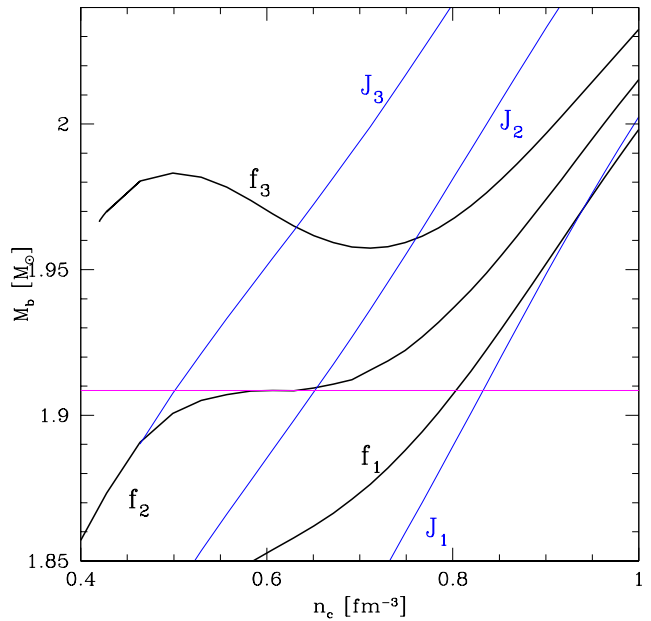}
}
\caption{Baryon mass {\it vs.} central baryon density for a schematic EOS softened by a mixed phase. The rotational frequency is fixed along black, thick curves. Blue (thin) lines correspond to the fixed total angular momentum.}
\label{fig:mbnc}
\end{figure}

In fig.~\ref{fig:mbnc} rotating configurations exhibit back-bending for frequencies $f>f_2$
and baryon masses larger than $\simeq 1.91\;\msun$ marked by a thin, red horizontal line. 
All configurations presented in fig.~\ref{fig:mbnc}
are dynamically stable, $(\partial \Mb/\partial n_c)_{J} > 0$.

\begin{figure}

\resizebox{\hsize}{!}{%
  \includegraphics{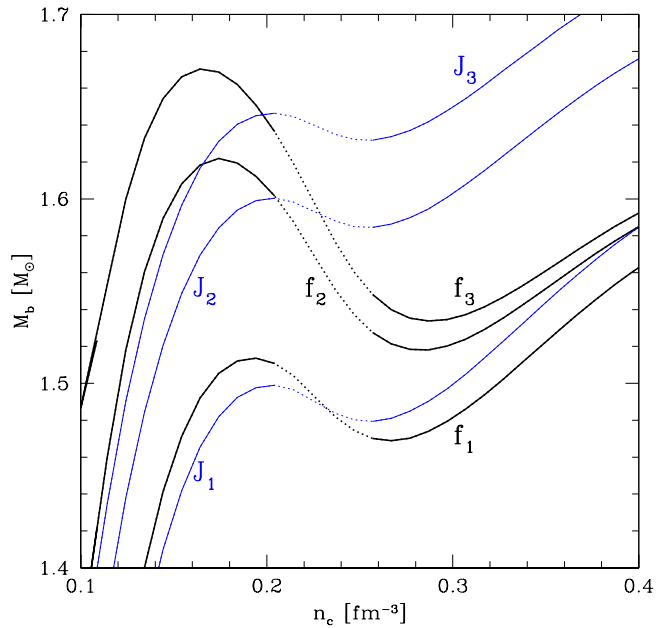}
}
\caption{Baryon mass {\it vs.} central density for the schematic EOS softened by a mixed phase. Strong softening 
results in the existence of unstable configurations (dotted lines).}
\label{fig:mbncin}
\end{figure}

In fig.~\ref{fig:mbncin} the case of strong softening is presented. There exists a region of
dynamically unstable configurations defined by the condition 
$(\partial \Mb/\partial n_c)_{J} < 0$.

The decreasing parts of $\Mb(n_c)$ relations, marked by dotted lines in fig.~\ref{fig:mbncin}, which correspond to unstable configurations define the instability strip separating 
two branches of stable, rotating configurations - one, less compact, with maximum mass
defined by the onset of instability due to the softening of EOS and 
the second one with maximum mass corresponding to the threshold for the 
collapse to black hole.
The analysis of the large set of EOS with softening through a mixed phase 
or first order phase transition at constant pressure indicates that the existence of these two
families does not depend on the rotational frequency (however, the width of the instability 
strip does depend on $f$). Equivalently the instability
strip starts at non-rotating configurations and continues up to the Keplerian 
limit - these two families are disjoint in the $M(R_{\rm eq})$ or $M(P_c)$ plane.

 The turning-point theorem is a {\it sufficient} condition.   
Recent numerical simulations by \cite{TakamiRY2011} study the onset of the 
dynamical instability for rotating stars and obtain it slightly below the maximum mass 
(at fixed angular momentum $J$). The relative difference of $M$ is of the order of $10^{-3}$ 
for rapidly rotating configurations; for most astrophysical purposes the turning-point 
theorem is therefore precise enough to locate the instability regions by means 
of stationary calculations.  

\subsection{First order phase transition and instability}
\label{sect:firstinstab}

For a continuous EOS ($P(\rho)$ with hyperonization or 
with a mixed-phase segment)  all macroscopic parameters  ($M$, $\Mb$, $J$) are
smooth functions of a central pressure $\pc$ or central density $\rho_{\rm c}$. As a consequence all
derivatives in eq.~(\ref{eq:M.MB.J.stab}) exist and extremal configurations corresponding to the 
onset of instability are defined by the vanishing of these derivatives:

\begin{equation}
\left({\partial M\over \partial \pc}\right)_{J} =0,\quad
\left({\partial \Mb \over \partial \pc}\right)_{J} =0,\quad
\left({\partial J\over \partial \pc}\right)_{\Mb} =0.
\label{eq:M.MB.J.stab0}
\end{equation}
However, if the EOS exhibits a first order phase transition ({\it e.g.} the Maxwell construction and phase transition to a new
phase at fixed pressure $\pne$, see sect.~\ref{pure.mixed}) the $P(\rho)$ relation is discontinuous
at $P=\pne$ with a density jump $\rhon \to \rhoe$. Let us stress that in the present section  we assume full
thermodynamic equilibrium (no metastability).

In this case an increase of the  central pressure due to evolutionary processes, 
resulting in  $P_{\rm c}>\pne$,  leads to the appearance of a small core of the dense phase at the center of a star with a
density jump ($\rhon \to \rhoe$) at the boundary separating the two phases. For a non-rotating star 
this density jump results in the non-continuous change of the derivatives of the main global 
parameters of a star  ($X=M, \Mb$) with respect to the central pressure \cite{zhs87}:
\begin{equation}
\left({\mathrm{d}X \over \mathrm{d}P_c}\right)_{\rm E}={(3-2\lambda+3\xn)(1+\xn) \over
(\lambda+3\xn)(\lambda+\xn)} \left({\mathrm{d}X \over \mathrm{d}P_c}\right)_{\rm N},
\label{eq:deriv}
\end{equation}
where $\lambda=\rhoe/\rhon$ is the density jump and $\xn\equiv {\pne/ \rhon c^2}$. The derivatives with
subscript N and E are taken, respectively, at a pressure infinitesimally smaller and larger than $\pne$. 
The direct consequence of eq.~(\ref{eq:deriv}) is the stability criterion for a stellar configurations
with a small core of the new, dense phase. The stability condition is given by the inequality
\cite{Seidov1971,Kaempfer1981}:
\begin{equation}
 \lambda<\lambda_{\rm crit}\equiv {3\over 2}(1+\xn).
 \label{eq:lcrit}
\end{equation}
We define a ``weak'' first order phase transition with a relatively small density jump for which the condition
in eq.~(\ref{eq:lcrit}) is fulfilled. This case is presented in fig.~\ref{fig:BM165Q} with a phase transition to quark matter
with $\lambda=1.2$. The softening of the EOS manifests itself as a sudden change of the slope of the $M(R)$ curve 
as a new phase of matter appears at the center.
For a ``strong'' first order phase transition ($\lambda>\lambda_{\rm crit}$) the derivatives $\mathrm{d}M/\mathrm{d}P_c$, 
${\mathrm{d}\Mb/\mathrm{d}P_c}$ change their sign at $P_c=\pne$ and stellar configurations with a small core of dense phase in the
center are dynamically unstable. 
The oscillatory mode (radial) which is in this case unstable has no counterpart for one-phase configurations. 
The main feature of this  oscillations is the flow of a matter through the pulsating boundary between two phases 
- the frequency of this mode is proportional to $\sqrt{3-2\lambda+3\xn}$, which directly means instability
for $\lambda>\lambda_{\rm crit}$  and collapse into a new configuration with a sizeable E-core \cite{HZS89}.


\section{Modeling of minicollapse}
\label{sect:minicoll}

\begin{figure}
\resizebox{\hsize}{!}{%
  \includegraphics{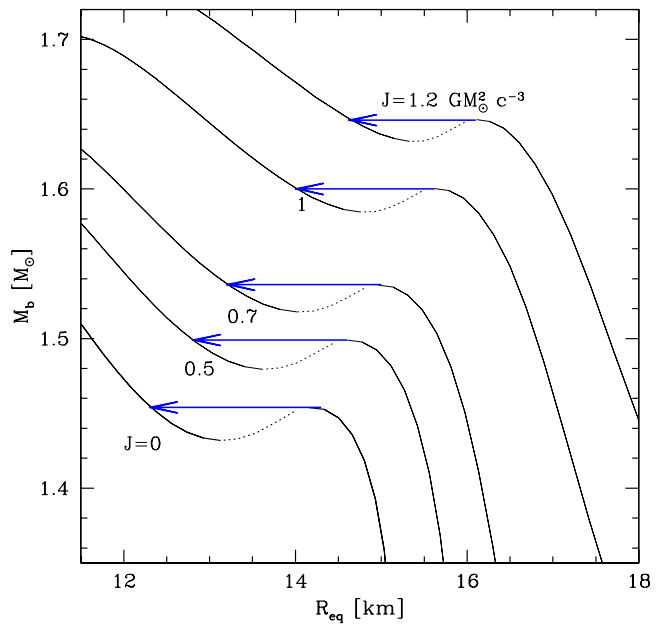}
}
\caption{Total baryon mass $M_{\rm b}$ {\it vs.} circumferential equatorial radius
$R_{\rm eq}$ for stationary non-rotating NS (thick line), and NS rotating at fixed angular momenta $J$. Dotted segments denote the unstable configurations.
The minicollapses are marked by arrows.}
\label{fig:mbrm}
\end{figure}
\subsection{Dynamics}
\label{sect:dynamics}
The existence of two disjoint families has important consequences for
the evolutionary tracks of isolated and accreting NS.
The loss of angular momentum moves an isolated NS leftwards along constant 
$\Mb$ lines on the $\Mb(R_{\rm eq})$ plane (fig.~\ref{fig:mbrm}). Once it reaches 
the instability strip, it collapses into a more compact counterpart 
with the same total angular momentum $J$ (arrows in fig.~\ref{fig:mbrm}). 
The dynamical properties 
of this minicollapse were studied using  general-relativistic (GR) numerical codes \cite{Dimmel2009,Abdikamalov2009}. The prefix mini is reflecting  the fact that the changes of
stellar parameters (radius, moment of inertia, frequency of rotation) associated with collapse under consideration are usually small. However, for a strong first order phase transition  the radius can shrink 
 by $\sim 10\%$ (see fig.17, 18). The distribution of a specific angular momentum in a collapsing star differs from 
 those in the  rigidly rotating initial and final configurations, because  a differential rotation profile develops during collapse. However, the
degree of differential rotation is found to be small and the assumption about 
 the conservation of the total angular momentum of the collapsing star $J$  seems to be  justified.  The energy release in the minicollapse turned out to depend weakly on $J$. However, it strongly depends on 
parameters of the EOS and of the phase transition itself. The formation of an E-core is followed by NS pulsations 
and the generation of gravitational waves, as described in more detail in the following subsections.
\subsection{Metastability, rotation,  and the total energy release}
\label{sect:meta-rot-energy}

The metastability of the N phase at the center of evolving NS
can be parametrized by the overpressure $\Delta P = P_{\rm crit}-\pne$ at which the E-phase
nucleates in the N-phase triggering a minicollapse from the initial configuration with central
pressure $P_c=P_{\rm crit}>\pne$. 

\begin{figure}
\resizebox{\hsize}{!}{%
  \includegraphics{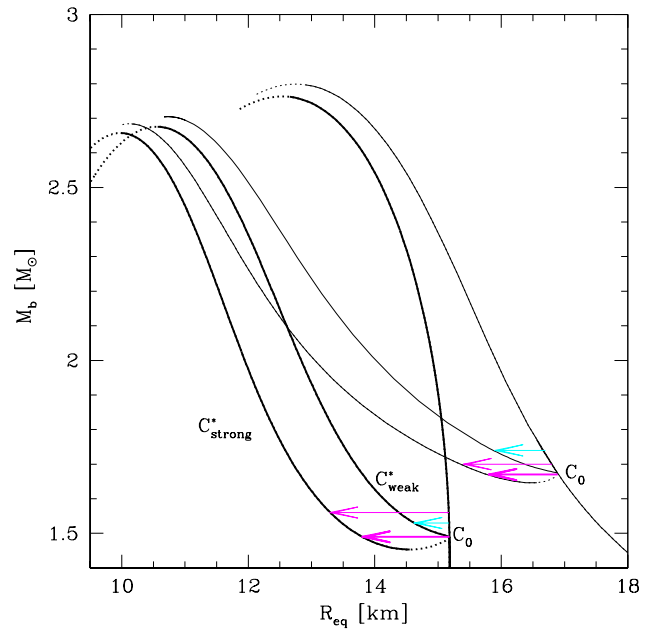}
}
\caption{Total baryon mass $M_{\rm b}$  {\it vs.} circumferential equatorial radius
$R_{\rm eq}$ for stationary non-rotating NS (thick lines) and NS rotating at a  fixed
angular momentum $J$, for two models of the phase transition weak and strong.  
Dotted segments correspond to unstable configurations.
The minicollapses are marked by arrows. The arrows above the
$C_0$ configuration correspond to a minicollapse from the metastable 
(overcompressed) phase $N$.}
\label{fig:mbrn}
\end{figure}

Mini-collapses for different $\Delta P$ are schematically
presented in fig.~\ref{fig:mbrn} in the case of weak and strong phase transition and for rotating
and non-rotating stars. Because for a strong phase transition two branches of stable configurations
are separated, the minicollapse is possible even without a overpressure $\Delta P=0$ (thick arrows in
fig.~\ref{fig:mbrn}) with a total energy release (difference) of the order of $\sim 10^{51}$ erg. For a weak first order
phase transition the overcompression and the existence of a metastable core is the only cause of  a
minicollapse and without overpressure the energy release is $\Delta E(\Delta P=0)=0$. An overcompression of the order
of $5-10\%$ leads to a $\Delta E\sim 10^{50}$~erg for a weak phase transition and to an increase of
$\Delta E$ by $\sim 10^{51}$~erg for a strong one \cite{ZdunikBHG2007,ZdunikBHG2008}. The dependence of the 
energy release on  $\Delta P$ is very weakly affected by the rotation rate and can be approximated
with a very high accuracy  by the values obtained for non-rotating stars (see \cite{ZdunikBHG2007,ZdunikBHG2008}).

%

\section{Astrophysical signatures of a minicollapse of neutron star}
\label{sect:mini-signatures}
 A phase transition in a NS core, that induces a dynamical minicollapse of NS, is associated with heating   of stellar interior, matter flow, and NS pulsations. It is important to remind, that the very essential  dynamical character of minicollapse is based on the assumption that the conversion of the matter into an exotic phase is of a detonation type.
We assume that the N$\longrightarrow$E  conversion front moves at supersonic speed,
 driving a shock wave in the N-envelope.

\subsection{Surface glowing, and a delayed re-brightening} 
\label{sect:glowing-rebright}
The dynamical and thermal effects  were studied in hydrodynamical simulations of spherically symmetric minicollapses induced by pion-condensation in hadronic matter in \cite{HaenselDenissov1990}, where the references to previous works can be found. The Newtonian approximation was used, and thermal effects,  such as heating and cooling, were considered. The initial temperature of the NS interior was assumed to be $10^8$~K. For an assumed model of pion-condensation, the total energy release $\Delta E$ was typically $\sim 10^{50}~{\rm erg}$ (notice that this quantity is strongly dependent on the phase-transition model). This energy  was liberated in $\sim 0.4$~ ms, and split into heating of the exotic core due to a latent heat and matter compression, and into heating of the N-matter outside the E-core due to the compression. The kinetic energy of the matter flow was mostly imparted  into the NS pulsations. The exotic (inner) core was heated to $\sim 10^{11}~$K, and the outer core and crust to some $10^{10}~$K. We expect that the shock wave strongly heated the NS surface \cite{Ramaty1980,Ramaty1981,Ellison1983} albeit this was not modelled in \cite{HaenselDenissov1990} due to a too low spatial grid resolution. The NS-core was rapidly cooled by neutrino radiation, so that after few hours most energy generated by the minicollapse  has been carried out by neutrinos. By that time NS pulsations have been damped. The heat content in the N-envelope, with cooling timescale orders of magnitude longer than for the E-core,  was diffusing to the NS surface, leading to its delayed brightening. This brightening was associated with surface  X-luminosity  $L_{\rm X}$ increasing during  30 yrs after minicollapse.  Then  $L_{\rm X}$ dropped by many orders of magnitude, on a much shorter timescale,  because the heat content of the N-envelope had been exhausted. 
\subsection{A gamma-ray burst?}
\label{sect:GRB}

A minicollapse  due to pion condensation was proposed to explain a famous very energetic  burst of gamma rays of 5th March 1979, coming from supernova remnant N49 in Large Magellanic Cloud \cite{Ramaty1980,Ramaty1981}. The initial gamma-ray pulse had a very short rise time of  $<0.4\;$ms, indicating a dynamical character of the burst mechanism. It was then followed by a pulsed decaying photon flux, suggesting NS rotation (8 s period was however a puzzle at that time).   A shock wave sent from the collapsed pion-condensed core propagated towards the NS surface, heating it to high temperature and allowing for an energetic burst of gamma rays. The  model was further elaborated in \cite{Ellison1983}. These authors used the  minicollapse parameters following the models of \cite{HaenselProszynski1982} as far as the energetics of the burst was concerned. The minicollapse scenario for the source of the 5th March 1979 event was further studied in \cite{Muto1990}. 

An essential progress in the gamma-ray astronomy,  started in 1990s,  led to a conclusion that the source of 
 5th March 1979 was  not a typical gamma-ray burster. It was actually an  exceptionally strong outburst from 
 a  soft-gamma ray repeater (SGR). SGRs belong to a subpopulation of magnetars, slowly rotating 
 (rotation period of several seconds, consistent with 8 s period of pulsations  in the tail of the 5th March 1979 burst)  and highly magnetised ($10^{14} - 10^{15}\;$G) NS. A SGR emits GRB at irregular intervals, from  hours to years and longer. The bursts from SGR  are powered by magnetic field, and they are triggered by crust-quakes and are associated with magnetic field annihilation and reconfiguration. Only seven SGR are known today, and they are obviously not related to minicollapse which occurs only once in a NS lifetime.  
 
Can a minicollapse due to a phase transition in NS core produce a class of regular non-repetitive GRBs that occur at cosmological distances larger than $100\;$Mpc, thousands of which  were detected since 1990s? Unfortunately, here the minicollapse and shock heating model faces a very basic problem, pointed out using detailed numerical simulations in  \cite{FryerWoosley1998}. Consider a  hot layer of relativistic plasma created by the shock wave at NS surface. It is an  element of a fireball which is a precursor of GRB. In order to produce a typical  GRB at cosmological distance such shock-produced fireball has not only to be sufficiently energetic ($E^{\rm f.b.}> 10^{51}~$erg) but should also 
contain a not too large amount of baryons ($M^{\rm f.b.}_{\rm b}<10^{-5}\;\msun$, so that the so-called Lorenz factor $\Gamma^{\rm f.b.} = E^{\rm f.b.}/M^{\rm f.b.}_{\rm b}c^2 > 100$ 
(see, e.g, \cite{MeszarosRees1993}).  In other words, the initial fireball should contain a nearly pure mixture of photons, neutrinos,  and $e^+e^-$ pairs that make it opaque: the contribution from the kinetic energy of baryons should be negligibly small.  Unfortunately, even for the most optimistic models for the shock-wave propagation (no neutrino losses, no photo-disintegration of nuclei)  the best one can get is $E^{\rm f.b.}=10^{46}\;$erg for  $\Gamma^{\rm f.b.} = 40$, which is only $10^{-5}$ of the required energy \cite{FryerWoosley1998}.

\subsection{A burst of gravitational waves}
\label{sect:GW}
Such astrophysical signature of a minicollapse was studied  in \cite{Dimmel2009,Abdikamalov2009}, which give also references to the previous work. The phase transitions considered were  associated with quark deconfinement  and kaon condensation. GR hydrodynamics was  used in the 3+1 formulation, but thermal effects were not included. Both studies were concentrated on dynamics of minicollapse of a rotating neutron star and on the potential importance of minicollapse as a detectable source of gravitational waves (GW).
The degree of differential rotation due to a minicollapse was found to be small.   Pulsations induced by a minicollapse,   when coupled to rotation,  break the  axial symmetry. This opens possibility of the GW  radiation in a burst (10 - 100 ms long), which if occurred at 10 kpc, could be detectable by the current second-generation interferometric detectors, like the Advanced LIGO (back online since September 2015), the Advanced Virgo (which will resume operations in the middle of 2016) \cite{Dimmel2009,Abdikamalov2009}, and the planned Einstein Telescope, a third-generation underground detector \cite{Punturo2010}.
\section{Effect of the crust formation scenario on the  $M - R$ relation}
\label{formation.crust}
It is usually assumed that the NS crust is composed of cold, catalyzed
matter. This is a good approximation when the outer layers of a NS are
formed at the birth in a stellar core collapse, when
$T>10^{10}\;$K allows for nuclear equilibrium in dense matter.
However, for a NS that passed through the long stage of accretion of
matter from a stellar companion in a binary system ({\it e.g.} NS recycled
to millisecond pulsars in low-mass X-ray binaries) the crust is
formed from the accreted layers of matter  and its composition is
widely different from that of catalyzed matter
\cite{HZ1990eos,HZ2008acc}. In view of these two possible formation
scenarios we have different EOS of NS crust, composed of catalyzed or accreted matter. 
The latter EOS is stiffer than the former, which   
results in a different thickness of the crust and different radii,
$R_{\rm acc}(M) > R_{\rm cat}(M)$ \cite{HZ1990eos,ZH2011crust}.

The formation of a fully accreted crust would take $\sim 10^{7}$ yrs for a
mean accretion rate $10^{-9}~\msun~{\rm yr}^{-1}$ typical of low-mass
X-ray binaries. For a $1.4\;\msun$ star and $R_{\rm cat}=12\;$km one gets $R_{\rm
acc}-R_{\rm cat}\simeq 100\;$m. The difference in the equatorial radius
grows faster than quadratically with the rotation frequency and
depends quite strongly on $R_{\rm cat}(f=0)$. For 716 Hz,
$1.4\;\msun$ and $R_{\rm cat}(f=0)=12\;$km the difference in
equatorial radii is 140 m \cite{ZH2011crust}. This means that the
effect of the formation scenario is significantly smaller than the
uncertainties in the measurements of $R$  (sect.~\ref{Obs}).

\section{Discussion and conclusions}
\label{conclusions}
In order to unveil the structure of dense matter of density up to
ten nuclear densities, we  confront theoretical models with measured NS
parameters. Theoretical models are legion. The discovery of two radio
pulsars of $2\;\msun$ resulted in an essential progress by putting
strong constraints on the EOS, but did not produce a satisfactory
answer to our fundamental question: do massive NS contain
exotic cores? Crucial for solving this problem are precise measurements
of NS radii with known masses. As for today, the
determinations of radii  are neither sufficiently precise nor
reliable to give a definite answer. Hopefully, the situation may change
in the future, thanks to the progress in X-ray astronomy. The
task is very challenging, and requires {\it e.g.} knowledge of distances
to NS with precision of the order of two percent. It is
regrettable that the LOFT mission will not fly in the near future,
because it was offering some very special opportunities for 
NS radii measurements {\cite{LOFT}. We can only hope that other
missions like NICER \cite{NICER} and Athena \cite{Athena+} will be successful 
in this respect. 

All NS rotate, and many of those which are good targets
for radius measurement are millisecond pulsars rotating at more
than 400 Hz. There is a rather strong interplay between the rotation and
the NS EOS, and therefore for both principal and practical
reasons it is advantageous and very often mandatory to use  accurate
GR formalism, with a consistent choice of space-time coordinates, to
calculate 2D hydrostatic equilibrium of rotating NS.
The slow-rotation approximation is not suitable for checking
stability, and does not allow the correct description of processes where
fulfilling conservation laws is crucial. Therefore we encourage to
use public domains precise 2D codes such as {\tt LORENE/nrotstar} or 
 \texttt{RNS}\footnote{\tt http://www.gravity.phys.uwm.edu/rns/} 
in the studies
that will eventually determine the true EOS of NS.

The use of precise 2D codes is particularly important for studying
phenomena associated with softening of the EOS due to the appearance
of the exotic phase. In particular, we explained this using
the example of the back-bending phenomenon and the loss/gain of
stability in the spin evolution of NS.

Since the discovery of $2\;\msun$ pulsars, there were numerous works
showing the possibility of the existence of  hyperonic matter models that
yield $M_{\rm max}>2\;\msun$. As we showed in the present review,
these dense matter models had a rather stiff pre-hyperon segment of
the EOS, resulting in rather large radii of stars with $M\sim
1.4\;\msun$, usually $R_{1.4}>12\;$km. We argue that this EOS feature 
is amplified by rotation, and leads also to a rather high minimum 
mass of rotating NS. 

A fully consistent calculation of the NS radius should in principle be performed for a unified EOS, 
where the crust and the outer layer of the core are described using the 
same nuclear model. Besides, `gluing' two different EOS based on different
nuclear models is, to a large extent, arbitrary and unphysical. 

The scenario for the formation of the crust (accretion in a close binary system 
or cooling down after the NS birth in supernova explosion) has a small effect
on the $M-R$ relation. The accreted crust is thicker by $\sim 100\;$m at
 $1.4\;\msun$. Rotation at 716 Hz increases this difference by some 50\%. 

The presence of a quark core in NS is not excluded by
observations of $2\;\msun$ pulsars. However, if a quark core in
such hybrid stars contains a sizeable fraction of the star's mass, quark
matter has to be stiff enough (sound speed $>0.6\;c$) and
deconfinement should occur at not too low density ($2.5n_0-3.5n_0$, 
and the density jump at the core edge should be small).
Finally, the last purely hadronic star should have a rather large
radius $>12~{\rm km}$ and a mass not much higher than $1.5\;\msun$
\cite{Zdunik13,AlfordHan2013}. Response to rotation is strong, with
$M_{\rm min}(716\;{\rm Hz})>1\;\msun$.

For a strong first order phase transition to quark matter (large density jump 
at the quark core edge) we obtain a 
family of very massive $\gtrsim 2\;\msun$ hybrid stars separated
from less compact massive hadron stars; in this way one finds
massive configurations of hadron and hybrid stars of the same
$M\simeq 2\;\msun$ but with a different structure and radius (hybrid and hadron
twins; the hadron twin has a significantly larger radius $>14~{\rm km}$
than the quark-core twin \cite{Benic2015twin}). We may expect that
the $M-R$ curve of these disjoint families will be quite sensitive to
rotation.

Generally, phase transitions in NS cores can have a strong
effect on the  $M - R$ relation for NS. Our discussion in
the case of the first-order phase transitions studied in the past
was actually very general. As we stressed,  one has to separate two
aspects of the phase transition ${\rm N} \longrightarrow {\rm E}$.
A crucial parameter for the $M - R$ relation for hydrostatic
equilibrium configurations is the energy density ratio at the phase
coexistence point $P=P_{\rm NE}$, $\lambda=\rho_{\rm E}/\rho_{\rm
N}$. Consider first non-rotating NS. For
$\lambda<\lambda_{\rm crit}=\frac{3}{2}\cdot (1+P_{\rm NE}/\rho_{\rm
E}c^2)$ equilibrium configurations form a continuous family with
$M_{\rm min}^{\rm stat}<M<M_{\rm max}^{\rm stat}$, with $M_{\rm
min}^{\rm stat}\simeq 0.1\;\msun$, and $M_{\rm max}^{\rm stat}>
2\;\msun$. This continuous character is conserved for configurations
rotating rigidly at $f$, with $M_{\rm max}^{\rm rot}$ increasing a
few percent at $716\;$Hz while $M_{\rm min}^{\rm stat}(716\;{\rm
Hz})\sim 0.7 - 1\; \msun$. For $\lambda>\lambda_{\rm crit}$ the
family of stable configurations of NS splits into two
disjoint families, separated by a segment of unstable configurations
(instability with respect to spherically symmetric perturbations)
that cannot exist in Nature. It has been checked that this
topological feature is  conserved for rotating 
configurations, with the instability being induced by axially-symmetric
perturbations. 

Transitions between two stable segments of the $M - R$ curve are
associated with NS minicollapse, with energy release weakly
depending on the angular momentum of the collapsing configuration, but
strongly depending on the degree of metastability
of the core undergoing the phase transition.

We reviewed various aspects of a minicollapse due to the formation of an exotic core. The most spectacular astrophysical signatures are associated with a minicollapse  induced by a N$\longrightarrow$E conversion of detonation type. We described the history of the famous  gamma-ray burst on March 5, 1979 and its initial explanation by a minicollapse in the core of  NS in N49 supernova remnant in Large Magellanic Cloud. The source of this burst turned out to be a soft-gamma repeater, a magnetar emitting  repetitively gamma and X-ray flares powered by the huge magnetic field. Finally, a positive message from recent numerical simulations of minicollapses  of rotating NS was that they could produce a burst of gravitational waves detectable in the Galaxy and its close vicinity by the network of Advanced Virgo and Advanced LIGO detectors, and in the future by the planned Einstein Telescope underground interferometric detector. 

{\it Acknowledgements} This work was partially supported by the Polish 
NCN grants no. 2014/13/B/ST9/02621 and 2013/01/ASPERA/ST9/00001.  
\appendix
 \section{Appendix}
\label{sec:app}
\begin{table}[h]
\center\begin{tabular}{lccc}
Object & Mass ($\msun$) & $f$ (Hz) & References\\
\hline
\multicolumn{4}{c}{Galactic NS+ WD}\\
\hline
J1804$-$2717	& 	$<1.73^\ast$			    	 & 	107.03 & \cite{TC99}\\
J1045$-$4509	& 	$<1.48^\ast$			 	     &  133.79 & \cite{TC99}\\
J1738+0333  	& 	$1.47^{+0.07}_{-0.08}$		 & 	170.94 & \cite{AK12}\\
J0437$-$4715	&	$1.44^{+0.07}_{-0.07}$ 	     & 	173.69 & \cite{RH15}\\
B1855+09    	&	$1.57^{+0.12}_{-0.11}$		 & 	186.49 & \cite{N08}\\
J1012+5307	    &	$1.64^{+0.22}_{-0.22}$	     & 	190.27 & \cite{CG98}\\
J1713+0747   	& 	$1.31^{+0.11}_{-0.11}$		 & 	218.81 & \cite{ZS15}\\
J2019+2425   	& 	$<1.51^\ast$					 & 	254.16 & \cite{NS01}\\
J0751+1807	    &	$1.26^{+0.14}_{-0.14}$		 & 	287.46 & \cite{NS05}\\
J1614$-$2230	&	$1.97^{+0.04}_{-0.04}$		 & 	317.38 & \cite{DP10}\\	
J1909$-$3744	& 	$1.47^{+0.03}_{-0.03}$		 & 	339.32 & \cite{RH15}\\
J0337+1715   	&	$1.4378^{+0.0013}_{-0.0013}$ & 	365.95 & \cite{RS14}\\
\hline
\multicolumn{4}{c}{NS+ WD in a GC}\\
\hline
J1748$-$2446I	& 	$1.91^{+0.02}_{-0.10}$ &	104.49 & \cite{KK13}\\
B1516+02B	& 	$2.08^{+0.19}_{-0.19}$ &	125.83 & \cite{FW08}\\
J0514$-$4002A	& 	$1.49^{+0.04}_{-0.27}$ &	200.38 & \cite{KK13}\\
J1910$-$5959A	& 	$1.33^{+0.11}_{-0.11}$ &	306.17 & \cite{CB12}\\
J0024$-$7204H	& 	$1.48^{+0.03}_{-0.06}$ &	311.49 & \cite{KK13}\\
\hline
\multicolumn{4}{c}{Galactic NS+MS}\\
\hline
J1903+0327	& 	$1.667^{+0.021}_{-0.021}$ &	465.14 & \cite{FB11}\\
J1023+0038	& 	$1.71^{+0.16}_{-0.16}$     &	592.42 & \cite{DA12}\\
\hline
\multicolumn{4}{c}{NS+NS in GC}\\
\hline
J1807$-$2500B	& 	$1.3655^{+0.0021}_{-0.0021}$ &	238.88 & \cite{LF12}\\
\hline
\end{tabular} 
\caption{Masses and spin frequency of radio MSP (with $f>100$ Hz). Masses are reported with a 1$\sigma$ uncertainty except for masses indicated by an $\ast$ symbol; $95\%$ confidence in this case.}
\label{tab:mf}
\end{table}

\end{document}